\documentclass[aps,pre,twocolumn,reprint,amssymb,amsmath,nobibnotes,nofootinbib
,superscriptaddress,floatfix]{revtex4}
\usepackage[dvips]{graphicx}
\usepackage[hidelinks]{hyperref}
\usepackage{epsfig}
\usepackage{framed}
\usepackage{color}
\usepackage{xcolor}
\usepackage{textcomp}
\usepackage{setspace}
\usepackage{morefloats}
\usepackage[T1]{fontenc}
\usepackage[utf8]{inputenc}
\usepackage{cleveref}
\usepackage{ulem}
\usepackage{comment}
\usepackage{float}
\usepackage{bm}

\begin{document}
\title{Polar flocks with discretized directions: the active clock model approaching the Vicsek model}
\author{Swarnajit Chatterjee}
\email{swarnajit.chatterjee@uni-saarland.de}
\affiliation{Center for Biophysics \& Department for Theoretical Physics, Saarland University, 66123 Saarbr{\"u}cken, Germany.}
\author{Matthieu Mangeat}
\email{mangeat@lusi.uni-sb.de}
\affiliation{Center for Biophysics \& Department for Theoretical Physics, Saarland University, 66123 Saarbr{\"u}cken, Germany.}
\author{Heiko Rieger}
\email{heiko.rieger@uni-saarland.de}
\affiliation{Center for Biophysics \& Department for Theoretical Physics, Saarland University, 66123 Saarbr{\"u}cken, Germany.}
\affiliation{INM – Leibniz Institute for New Materials, Campus D2 2, 66123 Saarbrücken, Germany.}

\begin{abstract}
We consider the off-lattice two-dimensional $q$-state active clock model (ACM) as a natural discretization of the Vicsek model (VM) describing flocking. The ACM consists of particles able to move in the plane in a discrete set of $q$ equidistant angular directions, as in the active Potts model (APM), with an alignment interaction inspired by the ferromagnetic equilibrium clock model. We find that for a small number of directions, the flocking transition of the ACM has the same phenomenology as the APM, including macrophase separation and reorientation transition. For a larger number of directions, the flocking transition in the ACM becomes equivalent to the one of the VM and displays microphase separation and only transverse bands, i.e. no re-orientation transition. Concomitantly also the transition of the $q\to\infty$ limit of the ACM, the active XY model (AXYM), is in the same universality class as the VM. We also construct a coarse-grained hydrodynamic description for the ACM and AXYM akin to the VM.
\end{abstract}

\maketitle


Active matter consists of particles that consume energy and convert it, for instance, into directed motion. Being manifestly out of equilibrium active matter systems display novel many-particle effects or collective phenomena, like flocking, motility induced phase separation, giant number fluctuations, active turbulence etc. New models have been developed in the last two decades to understand and unravel the physical principles governing active matter systems~\cite{AM-Reviews}. The paradigmatic model for collective motion of animal groups, like bird flocks, buffalo herds, fish schools, is the Vicsek model (VM)~\cite{VM}, in which particles moving with constant velocity align their direction of motion with the average direction of their neighbors. At low noise and large density the VM displays a flocking transition to collective motion in a common direction. Subsequent studies showed that the way in which noise and disorder are introduced into the system~\cite{Aldana,Peruani18}, range and type of the interactions~\cite{Peruani11,chate2008,Ginelli,Ginelli2} and alignment rules~\cite{Montagne,Mahault} influence the characteristics of pattern formation and the type of phase transition occurring in Vicsek-like models.

Even the nature of the flocking transition of the original VM was debated for a long time: originally thought to be continuous~\cite{VM} recent studies showed that it is discontinuous, reminiscent of a liquid-gas transition rather than a order-disorder transition~\cite{solon-vm}. In contrast to conventional first order phase transition scenarios, in which the system phase-separates macroscopically into a liquid and a gas phase in the coexistence region, the VM microphase separates into liquid bands of finite width moving coherently through the gas phase due to giant density fluctuations that break large liquid domains and arrest band coarsening~\cite{solon-vm}. Remarkably, such a microphase separation is absent in discretized versions of flocking models: the active Ising model (AIM)~\cite{AIM}, the $q$-state active Potts model (APM)~\cite{APM}, and an earlier version of the APM with volume exclusion effects~\cite{Peruani11} which shows surprisingly rich variety of self-organized patterns, all manifest macrophase separation in the coexistence region with only one liquid band moving in a gas background in a large aspect ratio rectangular geometry. In contrast to the VM, the APM displays additionally a reorientation transition from transversally moving bands for low particle velocities to longitudinally moving bands for high particles velocities~\cite{APM}.

A natural discretization of the VM (in 2d) is the 2d $q$-state active clock model (ACM), consisting of particles able to move in the plane in a discrete set of $q$ equidistant angular directions, as in the AIM or APM, with an alignment interaction inspired by the ferromagnetic equilibrium clock model~\cite{ECM}, approaching the ferromagnetic XY model in the limit $q\to\infty$~\cite{XYM}. Three questions arise in this context: 1) What is the nature of a putative flocking transition in the ACM for different values of the number of states $q$, regarding the fact that the equilibrium clock model has continuous BKT transition at a temperature $T_{BKT}$ into a quasi-long range ordered phase and for $q>4$ another transition at a temperature $T_{LRO}<T_{BKT}$ into a long-range ordered phase?~\cite{ECM2} 2) If the transition is first order, what are the characteristics of the coexistence region: microphase separation as in the VM or macrophase separation as in the AIM and APM? Do longitudinally moving bands exist? 3) Is the $q\to\infty$ limit of the ACM, the active XY model (AXYM) equivalent to the VM or does it remain macrophase separating as for finite $q$. 

In this letter we will answer these question and will show that for a small number of states, the flocking transition of the ACM has the same phenomenology as the APM \cite{APM}, including macrophase separation and reorientation transition. For a larger number of states, the flocking transition in the ACM becomes equivalent to the one of the VM and displays microphase separation and only transverse bands, i.e. no re-orientation transition. Concomitantly also the transition of the AXYM is in the same universality class as the one of the VM. The letter is organized as follows: first we define the ACM in detail, then we present our numerical results and our hydrodynamic theory, and finally we discuss the implication of our findings.

\medskip


{\bfseries Model.} -- The 2d $q$-state ACM consists of $N$ particles moving in a off-lattice rectangular domain of size $L_x \times L_y$ with periodic boundary conditions and with average particle density $\rho_0 = N / L_x L_y$. Since no mutual exclusion among the particles are considered, $\rho_0$ can assume values larger than $1$. Each particle carries a clock degree of freedom or angle $\theta\in \{0,2\pi/q,4\pi/q,\cdots, 2(q-1)\pi/q\}$, which also defines its preferred direction of motion in a biased diffusion. It can either jump to a new position or flip its angle. The hopping rate of a particle in state $\theta$ in the (discrete) direction $\phi$ is $W_{\rm hop}=D[1-\varepsilon/(q-1)]$ for $\phi \ne \theta$ and $W_{\rm hop}=D(1+\varepsilon)$ for $\phi = \theta$, where $D>0$ is the diffusion constant and $\varepsilon\in[0,q-1]$ is the bias, or ``velocity". Note that the total hopping rate is $W_{\rm hop}^{\rm tot}=qD$ and that $\varepsilon=0$ corresponds to unbiased diffusion and $\varepsilon=q-1$ to ballistic motion in the direction of the clock angle. If the hopping angle is $\phi$ and we denote the position of the $i^{\rm th}$ particle at time $t$ by ${\bf x_i}(t)$, then its position in the next time step is ${\bf x_i}(t) + {\bf e_\phi}$, where ${\bf e_\phi}$ is the unit vector in $\phi$-direction. The flipping rate from $\theta_i = \theta$ to $\theta_i=\theta'$ is derived via detailed balance from a local clock Hamiltonian
\begin{equation}
\label{Hclock}
H_i=-\frac{J}{2\rho_i} \sum_{k\ne l, k,l\in{\cal N}_i} \cos(\theta_k-\theta_l) \; ,
\end{equation}
where $J$ is the ferromagnetic coupling constant and $\rho_i$ is the number of particles within its neighborhood ${\cal N}_i= \{j\;{\rm with}\;|{\bf x}_i - {\bf x}_j| \leqslant 1\}$:
\begin{equation}
\label{Wflip}
W_{\rm flip} = \gamma \exp \left\{\frac{\beta J}{\rho_i}\left[{\bf m_i} \cdot( {\bf e_{\theta '}} - {\bf e_{\theta}}) + 1 - \cos(\theta-\theta')\right] \right\}\;,
\end{equation}
where ${\bf m_i} = \sum_{j\in{\cal N}_i} (\cos \theta_j,\sin \theta_j )$ is the local magnetization and $\gamma$ is a constant. The origin of the term $1 - \cos(\theta-\theta')$ in Eq.~\eqref{Wflip} is the absence of self-interaction in the clock Hamiltonian in Eq.~\eqref{Hclock} (see \cite{SI} for a detailed explanation). Although phenomenologically ACM is very similar to the APM~\cite{APM}, the Kronecker delta ``Potts'' interaction in the APM has been replaced by the cosine ``clock'' interaction in the ACM motivated by the $q \to \infty$ limit and whether one recovers the VM in that limit and here the ACM is clearly better suited than the APM.

For $q=4$ and $6$, one can define the ACM on square and triangular lattices, respectively, and identify the $q$ different directions of motion to the $q$ nearest neighbors. We analyzed these lattice versions, too, and obtained qualitatively identical results as those reported below, see~\cite{SI}, but here we restrict ourselves to the off-lattice version which allows a straightforward $q\to\infty$ limit (AXYM) and is also closer to the original VM. In the limit $q\to\infty$ the rescaled quantities $\overline{D} = qD$ and $\overline{\varepsilon} = \varepsilon/(q-1)$ have to stay finite and the angles become continuous with $\theta \in [0,2\pi]$. The jump rate of a particle in state $\theta$ in the (continuous) direction $\phi$ becomes $W_{\rm hop}=\overline{D}(1-\overline{\varepsilon})$ for $\phi \in [0,2\pi]$ and $W_{\rm hop} = \overline{D}\overline{\varepsilon}$ for $\phi = \theta$.

We performed Monte Carlo simulations of the $q$-state ACM and the AXYM, which evolve in discrete time steps of length $\Delta t$. In each time-step $N$ (=number of particles) single particle updates are performed, one of which consists in choosing randomly a particle which then either updates its spin state to $\theta' \ne \theta$ chosen randomly with probability $p_{\rm flip} = W_{\rm flip} \Delta t$, or hops to one of the $q$ directions with probability $p_{\rm hop} = \overline{D} \Delta t$: in a random direction with probability $(1-\overline{\varepsilon})p_{\rm hop}$ or in the direction $\theta$ with probability $\overline{\varepsilon}p_{\rm hop}$. The probability that nothing happens during this single particle update is $p_{\rm wait} = 1 - p_{\rm flip} - p_{\rm hop}$. An expression for $\Delta t$ can be chosen to minimize $p_{\rm wait}$: $\Delta t = [\overline{D} + \exp(2\beta J)]^{-1}$. This is a hybrid dynamics combining Monte Carlo and a real-time dynamics previously used in the simulations of the AIM~\cite{AIM} and the APM~\cite{APM}. Without any loss of generality, we can take $\overline{D}=1$, $J=1$ and $\gamma=1$.

We consider mainly a rectangular domain with a large aspect ratio with $L_x=400$ and $L_y=50$ for the computation of the phase diagram and other quantities. $L_x=800$ with varying $L_y$ are considered for the snapshots presented. Simulations are performed for three control parameters: the noise is regulated by $\beta=1/T$, $\rho_0=N/L_xL_y$ defines the average particle density, and $\overline{\varepsilon}$, the self-propulsion parameter, dictates the effective velocity of the particles. The initial homogeneous system is prepared by assigning random initial position $(x_i,y_i)$ and orientation $\theta_i$ to each particle and then we let the system evolve under various control parameters for $t_{\rm eq} = 10^5\Delta t$ to reach the steady-state. Following this, measurements are carried out with a maximum simulation time $t_{\rm max} = 20 t_{\rm eq}$.

\medskip


\begin{figure}
\centering
\includegraphics[width=\columnwidth]{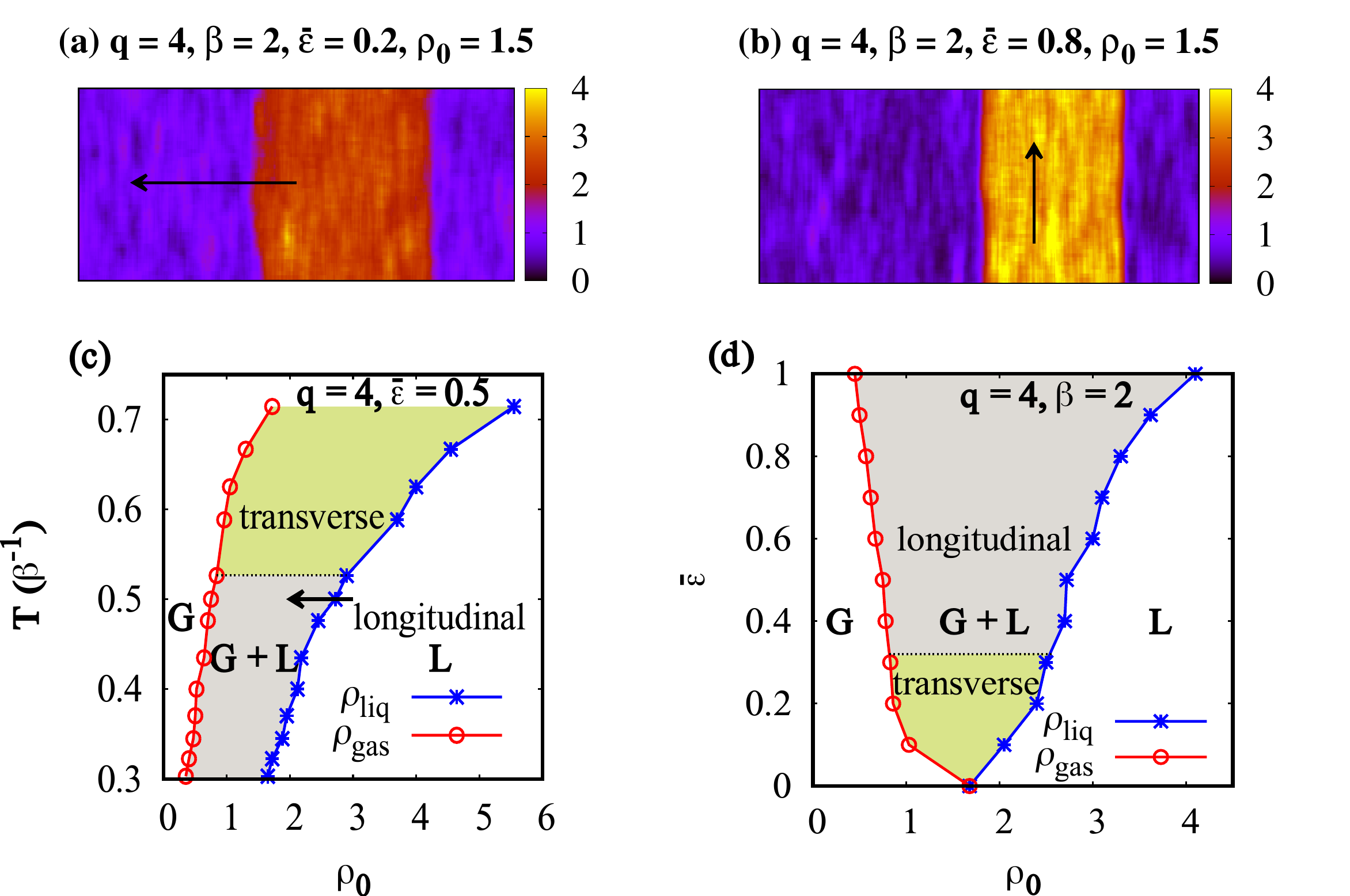}
\caption{(Color online) (a)--(b)~Stationary density profiles for $q=4$, $\beta=2$ and $\rho_0 = 1.5$ in a $400 \times 50$ domain showing the bulk phase-separation and reorientation transition from (a)~transverse band motion for $\overline{\varepsilon} = 0.2$ to (b)~longitudinal band motion for $\overline{\varepsilon} = 0.8$. The colorbar represents the particle density. (c)~Temperature-density ($T$-$\rho_0$) phase diagram for $\overline{\varepsilon}=0.5$, and (d)~Velocity-density ($\overline{\varepsilon}$-$\rho_0$) phase diagram for $\beta=2$ of the $q=4$-state ACM. The reorientation transition happens at $\beta=1.9$ and $\overline{\varepsilon}=0.32$, respectively.}
\label{figure_Q4}
\end{figure}

{\bfseries Phase diagrams and coexistence region.} -- In Fig.~\ref{figure_Q4}(a) and Fig.~\ref{figure_Q4}(b), we show stationary density profiles for the $q=4$-state ACM on a $400 \times 50$ rectangular domain for fixed $\beta=2$ and $\rho_0=1.5$ but for different bias: (a) $\overline{\varepsilon}=0.2$ and (b) $\overline{\varepsilon}=0.8$. As observed before in the VM~\cite{VM,solon-vm}, the AIM~\cite{AIM}, and the $q$-state APM~\cite{APM}, the transition from a homogeneous gas phase to a polar liquid phase occurs through a liquid-gas coexistence phase, where a single band of polar liquid propagates on a disordered gaseous background. Akin to the $4$-state APM~\cite{APM}, the band moves (a) transversally at small bias (velocity) $\overline{\varepsilon}=0.2$ and (b) longitudinal with respect to the band direction at larger bias $\overline{\varepsilon} = 0.8$. The coexistence phase in both figures shows a fully phase-separated density profile with a single {\itshape macroscopic} liquid domain as observed previously in the context of lattice flocking models~\cite{AIM,APM}. In Fig.~\ref{figure_Q4}(c) and Fig.~\ref{figure_Q4}(d), we display the temperature-density ($T$-$\rho_0$) phase diagram for $\overline{\varepsilon}=0.5$ and the velocity-density ($\overline{\varepsilon}$-$\rho_0$) phase diagram for $\beta=2$, respectively. The liquid and gas binodals $\rho_{\rm liq}$ and $\rho_{\rm gas}$, which segregate the gas-liquid (G+L) coexistence phase from the two homogeneous phases, liquid (L) and gas (G), are extracted from the time-averaged phase-separated density profiles. Reported for the first time in the context of the APM~\cite{APM}, we observe a similar reorientation transition of the coexistence phase from transverse band motion at low velocities and high temperatures to longitudinal lane formation at high velocities and low temperatures for $q=4$. The physical origin of this reorientation transition, as argued in Ref.~\cite{APM} with equivalent hopping rules, is the decrease of the transverse diffusion constant for large velocities, stabilizing the longitudinal lane formation. The reorientation transition occurs at $\beta=1.9$ (c) and $\overline{\varepsilon}=0.32$ (d), where the black dotted lines delimit the two co-existing phase domains which are further marked by two distinct colors: grey for longitudinal lane motion and yellow for transverse band motion. We have obtained similar results from the numerical simulations of the 4-state ACM on a square lattice~\cite{SI}.

\begin{figure}
\centering
\includegraphics[width=\columnwidth]{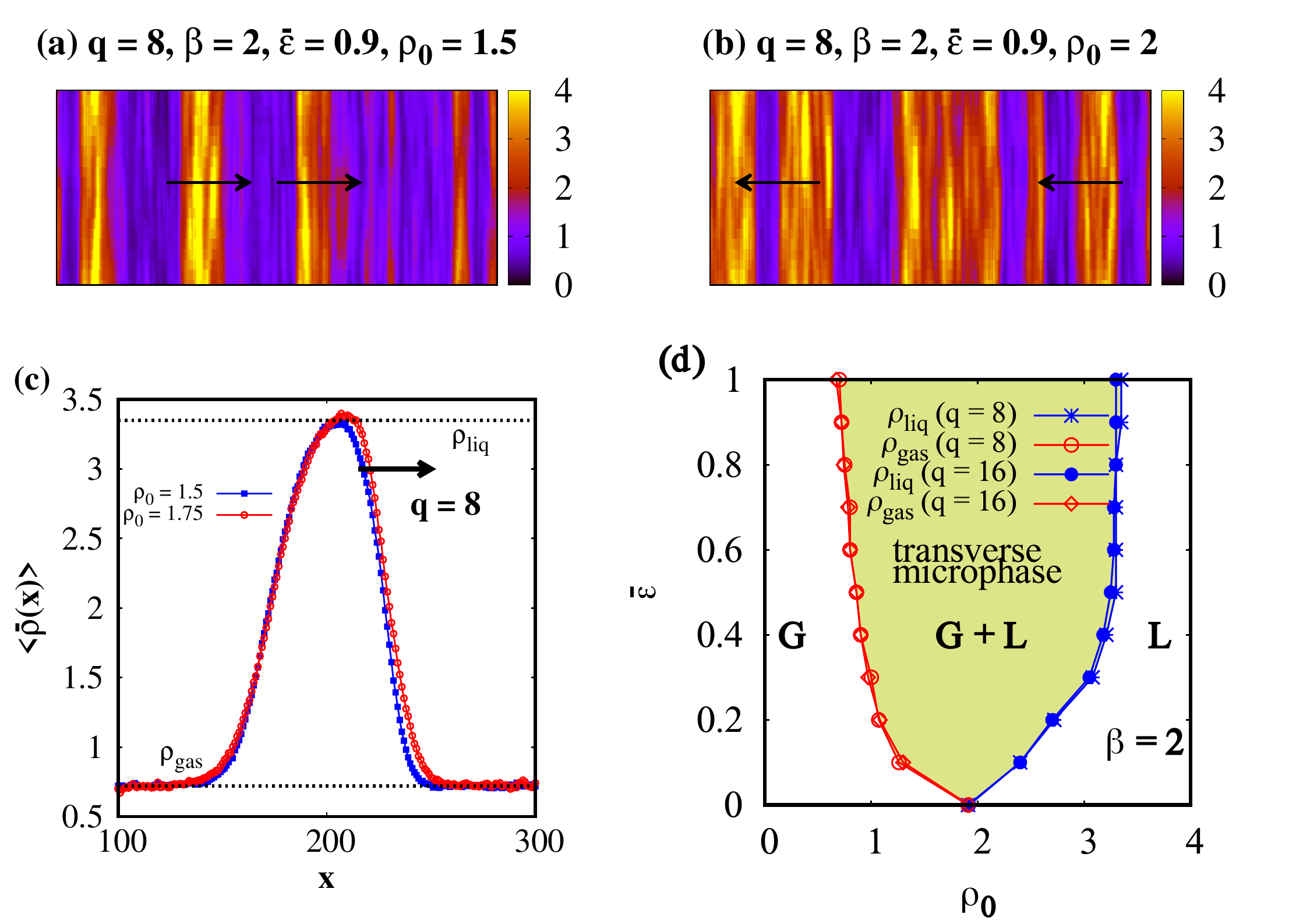}
\caption{(Color online) (a)--(b)~Stationary density profiles for the $q=8$-state ACM, $\beta=2$ and $\overline{\varepsilon} = 0.9$ in a $800 \times 20$ domain showing transversely moving microphase-separated bands for (a) $\rho_0=1.5$ and (b) $\rho_0=2$. The colorbar represents the particle density. (c) Time-averaged density profiles for $\rho_0 \in \{1.5,1.75\}$ defining the two binodals $\rho_{\rm liq}$ and $\rho_{\rm gas}$. (d)~Velocity-density ($\overline{\varepsilon}$-$\rho_0$) phase diagrams for $q=8$ and $q=16$-state ACM obtained for $\beta=2$ showing only transverse band motion.}
\label{figure_Q8}
\end{figure}

In Fig.~\ref{figure_Q8}(a) and Fig.~\ref{figure_Q8}(b), we show stationary density profiles for the $q=8$-state ACM on a $800 \times 20$ rectangular domain for $\beta=2$, $\overline{\varepsilon}=0.9$ and (a) $\rho_0=1.5$ and (b) $\rho_0=2$. A {\itshape microphase} separation of the coexistence region, where periodically arranged ordered liquid bands move in the same direction in a gaseous background, is observed. The microphase-separated traveling bands are transverse in nature as observed first in the VM~\cite{VM}. In a {\itshape microphase} separation, the traveling bands are not fully phase-separated and as established in Ref.~\cite{solon-vm}, one crucial characteristic of this {\itshape microphase} separation is that the band number $n_b$ increases with the density $\rho_0$ as observed in Figs.~\ref{figure_Q8}(a)-(b). Time-averaged density profiles of the liquid-gas coexistence phase are shown in Fig.~\ref{figure_Q8}(c) which suggest that the width of the polar liquid band does not increase significantly with the average density $\rho_0$ (see \cite{SI} for the algorithm which has been used to obtain the time-averaged profiles). It is well known that the band width does not affect the liquid ($\rho_{\rm liq}$) and the gas ($\rho_{\rm gas}$) binodals and we use this property to extract the relevant phase diagrams. In Fig.~\ref{figure_Q8}(d), we represent the velocity-density ($\overline{\varepsilon}$-$\rho_0$) phase diagrams for $\beta=2$ and for $q=8$ and $q=16$-state ACM. The two diagrams are very similar both qualitatively and quantitatively and implying a similar physical picture of the $q$-state ACM for $q \geqslant 8$. The corresponding coexistence domain of the $q=8$ and 16-state ACM is completely described by transversely traveling {\itshape microphase} separated bands. Although a reorientation transition occurs for $q=6$-state ACM when simulated on a triangular lattice~\cite{SI}, we do not observe any reorientation transition for off-lattice simulations for $q \geqslant 6$ at large bias $\overline{\varepsilon}$ as observed for $q=4$. 

\begin{figure}
\centering
\includegraphics[width=\columnwidth]{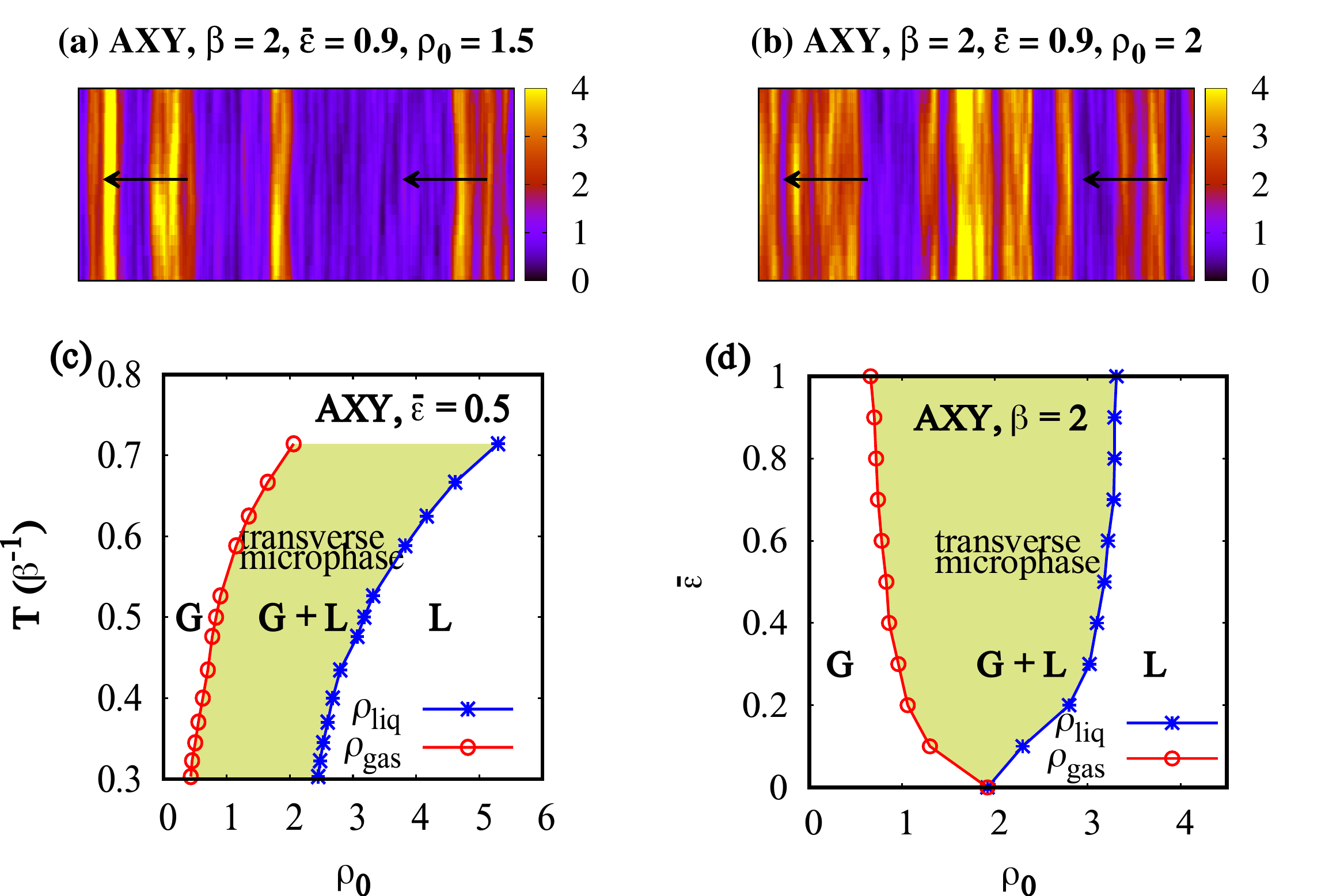}
\caption{(Color online) (a)--(b)~Stationary density profiles for the AXYM, $\beta=2$ and $\overline{\varepsilon} = 0.9$ in a $800 \times 20$ domain showing transversely moving microphase-separated bands for (a) $\rho_0=1.5$ and (b) $\rho_0=2$. The colorbar represents the particle density. (c) Temperature-density ($T$-$\rho_0$) phase diagram for $\overline{\varepsilon}=0.5$, and (d)~Velocity-density ($\overline{\varepsilon}$-$\rho_0$) phase diagram for $\beta=2$ for the AXYM showing only transverse band motion.}
\label{figure_AXY}
\end{figure}

In Fig.~\ref{figure_AXY}(a) and Fig.~\ref{figure_AXY}(b), we present the stationary density profiles in the coexistence phase of the AXYM (i.e. for $q = \infty$) on a $800 \times 20$ rectangular domain and for $\beta = 2$, $\overline{\varepsilon}=0.9$ and (a) $\rho_0=1.5$ and (b) $\rho_0=2$. The AXYM and VM possess the same $O(2)$ rotational symmetry but differ in their flipping and hopping rules. Nevertheless, we observe a {\itshape microphase} separation in the coexistence regime like in the VM~\cite{solon-vm} where the traveling bands are moving in the same direction and $n_b$ is increasing with $\rho_0$. The temperature-density ($T$-$\rho_0$) and the velocity-density ($\overline{\varepsilon}$-$\rho_0$) phase diagrams are shown in Fig.~\ref{figure_AXY}(c) for $\overline{\varepsilon}=0.5$ and Fig.~\ref{figure_AXY}(d) for $\beta=2$, respectively. We do not observe the reorientation transition akin to the observation made for $q=8$ and $q=16$-state ACM. The velocity-density ($\overline{\varepsilon}$-$\rho_0$) phase diagram is also identical to Fig.~\ref{figure_Q8}(d) both qualitatively and quantitatively and thus minimizing the statistical errors in the calculations of the binodals, these three diagrams can be merged in a single diagram which signifies that the system behaves similarly for large number of directions or large $q$ values. Moreover, for the $q$-state ACM and the AXYM we recover the characteristic velocity-density ($\overline{\varepsilon}$-$\rho_0$) phase diagram observed in other discrete flocking models~\cite{AIM,APM}. However, the nature of the $\bar{\varepsilon} \to 0$ transition of the AXYM is different from the VM. The density at which the gas and liquid binodals intersect at $\overline{\varepsilon}=0$ is finite ($\rho^*=1.95$) for the AXYM whereas it is infinite for the VM, as argued in \cite{solon-vm}.

\begin{figure}
\centering
\includegraphics[width=\columnwidth]{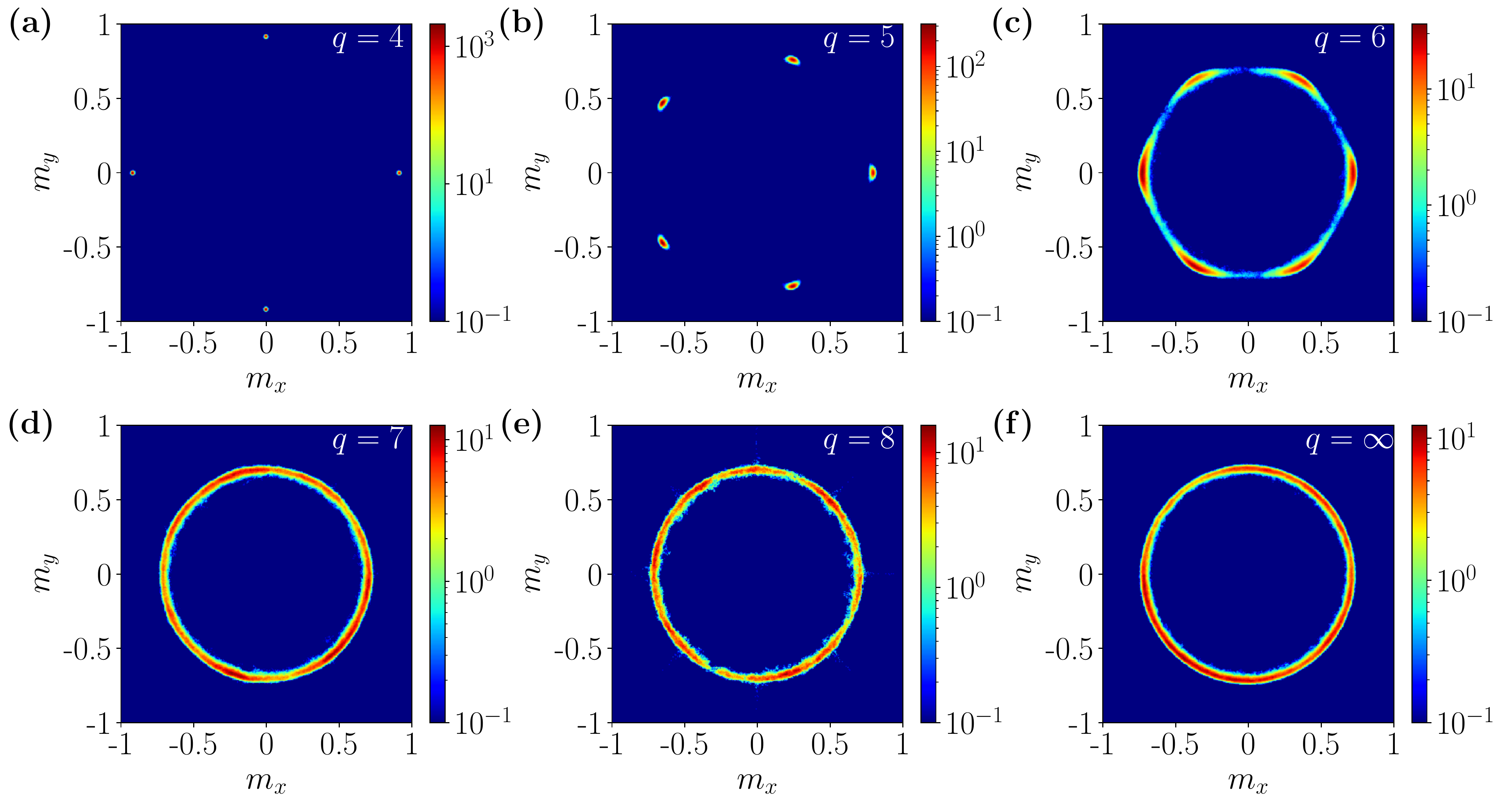}
\caption{(color online) Zero-activity limit of the order parameter distributions of the q-state ACM and AXYM ($q \to \infty$) in the liquid phase. Parameters: $L=50$, $\beta=2$, $\bar{\varepsilon}=0$, and $\rho_0=3$.}
\label{figure_e0}
\end{figure}

{\bfseries Zero activity limit ($\bar{\varepsilon}=0$).} -- This limit is denoted as the Brownian clock model, reminiscent of the Brownian Potts model studied in~\cite{HR2022}. We observe an order-disorder phase transition without a coexistence region, as observed for the AIM~\cite{AIM} and the APM~\cite{APM}. In Fig.~\ref{figure_e0}, we show the distribution of the order parameter $\mathbf{m}=(m_x,m_y)$ with $m_x=\sum_{i=1}^N \cos \theta_i$ and $m_y=\sum_{i=1}^N \sin \theta_i$ in the ordered phase for $\beta=2$, $\bar{\varepsilon}=0$, and $\rho_0=3$ simulated on a square domain of system size $L=50$ and averaged over time and several initial configurations. In Fig.~\ref{figure_e0}(a) and (b), we observe a well defined long-range ordered phase (LRO) for $q=4$ and $q=5$, respectively, where the distributions manifest $q$ isolated spots (pinned orientations) corresponding to the $q$-fold degeneracy of the ordered liquid phase with equal probability. In Fig.~\ref{figure_e0}(c)--(f), one observes for $q \geqslant 6$, ringlike distributions (unpinned orientations) signifying the Kosterlitz-Thouless (KT) type phase or the quasi-long range ordered (QLRO) phase, where spin waves and vortices arrange the spin vectors. For discrete $q$ values, a LRO phase can be observed at large densities~\cite{SI} or small temperatures, which is not the case for the AXYM.

In the AXYM, particles can diffuse along any random direction with hopping rate $\bar{D}$, whereas, the VM reduces to the two-dimensional XY model at the zero velocity limit (with immobile particles). Although it has been shown for the Brownian Potts model~\cite{HR2022} that diffusion can change the nature of transition, the diffusive motion of the particles in the AXYM do not change the structure of the corresponding field theory compared to non-motile particles in the VM. Therefore, the Mermin-Wagner theorem is still applicable even though this system is driven out of equilibrium and the ordered phase we observe is QLRO in nature, akin to the XY model in two-dimension. The problem of diffusively moving spins, along with similar arguments, has also been studied explicitly in Ref.~\cite{Peruani2016} in the context of active phase oscillators with $O(2)$ symmetry, where QLRO is reported for normal diffusion of oscillators whereas, super-diffusive motion is needed in order to obtain long-range order in two dimensions.

\begin{figure}
\centering
\includegraphics[width=\columnwidth]{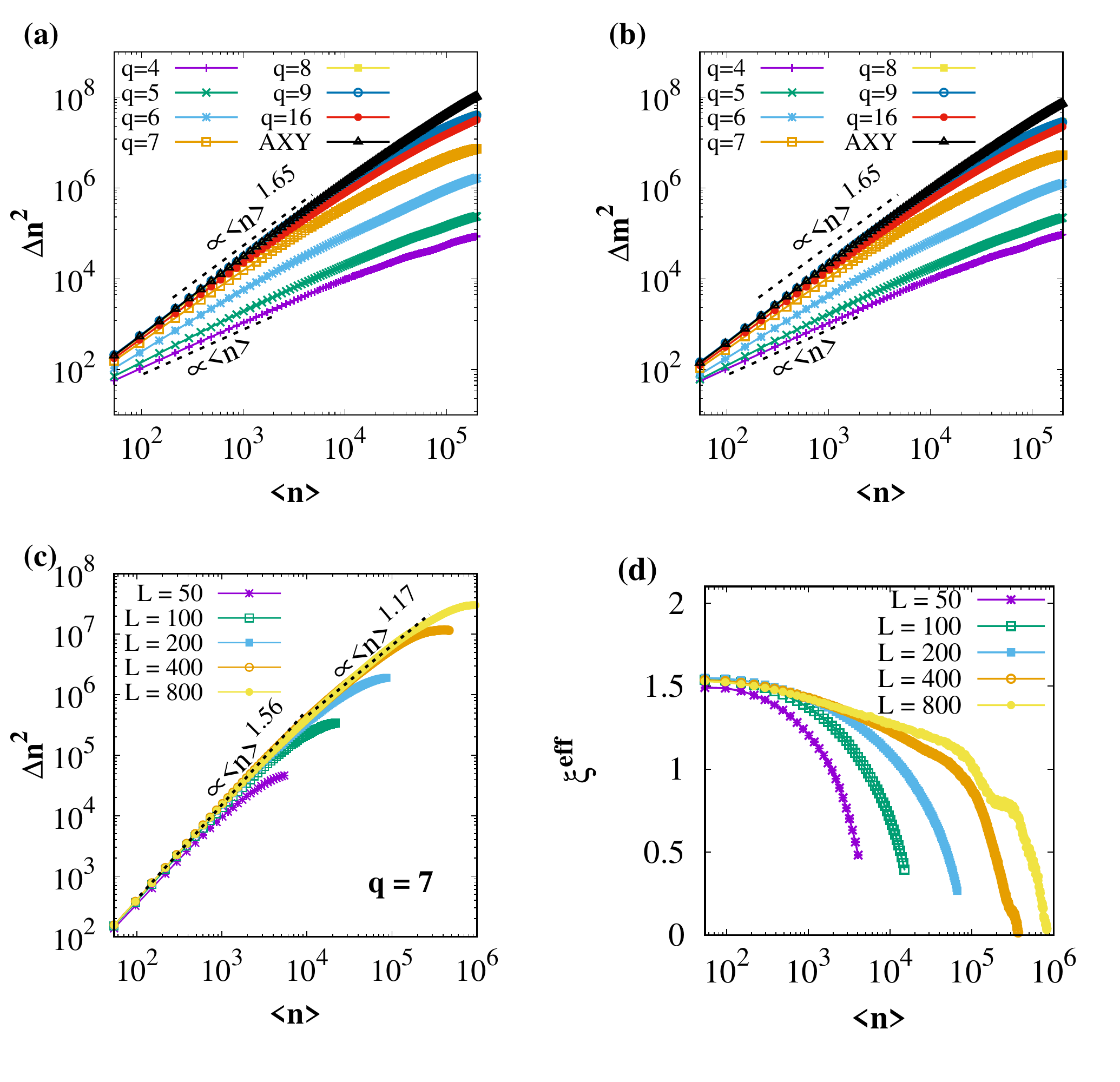}
\caption{(Color online) (a)--(b)~Number fluctuations $\Delta n^2=\langle n^2 \rangle-\langle n \rangle^2$ and magnetization fluctuations $\Delta m^2=\langle m^2 \rangle-\langle m \rangle^2$ versus average particle number $\langle n \rangle$ for several $q$ values in a $400 \times 400$ domain. (c)~Number fluctuations $\Delta n^2$ vs $\langle n \rangle$ as a function of various system sizes for $q=7$. (d) Effective exponent $\xi^{\rm eff}$ for the data plotted in (c). Parameters: $\beta=2$, $\bar{\varepsilon}=0.9$, and $\rho_0=6$.}
\label{figure_fluctuations}
\end{figure} 

{\bfseries Number fluctuations.} -- In Fig.~\ref{figure_fluctuations}(a)--(b), we show respectively the number fluctuations $\Delta n^2=\langle n^2 \rangle-\langle n \rangle^2$ and the magnetization fluctuations $\Delta m^2=\langle m^2 \rangle-\langle m \rangle^2$ for various $q$ values against the average particle number $\langle n \rangle$. $n$ and $m$ are respectively the number of particles and the magnetization in boxes of different sizes $\ell$ included in a $400 \times 400$ domain (with $\ell \leqslant 200$), with $\langle n \rangle=\rho_0\ell^2$. The data are for the liquid phase where $\beta=2$, $\overline{\varepsilon}=0.9$, and $\rho_0=6$. As shown in Table~\ref{table_exponents}, both the fluctuations behave like $\langle n \rangle^\xi$ with the fluctuation exponent $\xi \simeq \xi_n \simeq \xi_m$ increasing with $q$, from $\xi \simeq 1$ for $q=4$ to $\xi \simeq 1.65$ for large $q$ (and saturates for $q \geqslant 8$). Consequently the number and magnetization fluctuations show a transition from uniform fluctuations for small $q$ to giant fluctuations at larger $q$ as they have been observed in the VM~\cite{solon-vm}. Although the existence of giant number fluctuations (GNF) were shown in Vicsek-like self-propelled particle models~\cite{chate2008, Ginelli}, the connection between GNF and micro-/macro- phase separation was first hypothesized in Ref.~\cite{solon-vm} in the context of VM where it has been argued that GNF ($\xi_n \simeq 1.6$) break large bulk liquid domains and consequently produce smectic like microphase state in the coexistence regime whereas the system undergoes bulk phase separation when the density fluctuations are normal ($\xi_n \simeq 1$)~\cite{AIM}. In the ACM therefore, these GNF for large $q$ might be responsible for the microphase separation in the coexistence regime as shown in Fig.~\ref{figure_Q8}(a)--(b) and Fig.~\ref{figure_AXY}(a)--(b) for $q=8$ and $q=\infty$, respectively. Nevertheless one should stress that a causal relation between the existence of GNF in the ordered phase and the existence of micro-phase separation in the coexistence phase, as conjectured in~\cite{solon-vm}, is still hypothetical and remains an interesting open question. A comparison between Fig.~\ref{figure_fluctuations}(a) and Fig.~\ref{figure_fluctuations}(b) clearly reveals that GNF correspond to giant magnetization fluctuations or phase fluctuations which physically signifies weaker phase-ordering, and uniform number fluctuations correspond to smaller magnetization fluctuations which physically signifies stronger phase-ordering. 

The finite size effect on the number fluctuations for $q=7$ is shown in Fig.~\ref{figure_fluctuations}(c) where the data can be fitted to two different power-law regimes (consider the largest system size $L=800$) and one can extract: (i) an exponent of 1.56 in the interval $[10^2,10^3]$ and (ii) an exponent of 1.17 in the interval $[10^4,10^5]$. These exponents along with the exponents tabulated in Table~\ref{table_exponents} have been obtained by fitting the data to a power-law and since what one obtains depends on the x-range to which one restricts the fits, we have a look at the log-log slope or the corresponding effective exponent $\xi^{\rm eff}=d[\ln(\Delta n^2)]/d[\ln \langle n \rangle]$ plotted in Fig.~\ref{figure_fluctuations}(d) (see~\cite{SI} for the effective exponents corresponding to Fig.~\ref{figure_fluctuations}(a)--(b)). The plot shows a ``plateau'' around the first exponent $\xi \simeq 1.56$ but we observe no such ``plateau'' around the second exponent $\xi \simeq 1.17$. Therefore, on the basis of our data, even for the largest system size, one cannot predict an asymptotic value of the effective exponent which might suggests a crossover from giant to conventional number fluctuations. Note that $\xi^{\rm eff}$ must decrease with increasing $\langle n \rangle$ when $\langle n \rangle$ approaches the total number of particles in the system and becomes smaller than 1 due to the finite-size cut-off at $\langle n \rangle=N=\rho_0L^2$, where $\Delta n^2$ vanishes. 

\begin{table}

\begin{center}
\begin{tabular}{ |c|c|c|c|c|c|c|c|c| } 
\hline
$q$ & $4$ & $5$ & $6$ & $7$ & $8$ & $16$ & $\infty$ \\
\hline
$\xi_n$ & $1.04$ & $1.08$ & $1.36$ & $1.56$ & $1.62$ & $1.62$ & $1.65$ \\
\hline
$\xi_m$ & $1.06$ & $1.09$ & $1.37$ & $1.57$ & $1.63$ & $1.62$ & $1.65$ \\
\hline
\end{tabular}
\caption{Number fluctuation exponents $\xi_n$ and magnetization fluctuation exponent $\xi_m$ for several values of $q$, reported from Fig.~\ref{figure_fluctuations}. The typical error on the fluctuation exponents is $0.01$. \label{table_exponents}}
\end{center}

\end{table}

\medskip


{\bfseries Hydrodynamic description.} -- Next, we derive the main equations for the hydrodynamic continuum theory. From the microscopic hopping and flipping rates of the $q$-state ACM, we derive the master equation for the probability density function $n({\bf x},\theta;t)$ for a particle to be at the position ${\bf x}$ and in the spin-state $\theta$ at the time~\cite{SI}. We only keep the first-order terms in the $|{\bf m}_i| \ll \rho_i$ expansion in the flipping rate~\eqref{Wflip}. In the large system size limit $L \gg 1$, the hydrodynamic equation can be derived for the density $\rho({\bf x};t) = \int d\theta n({\bf x},\theta;t)$ and the magnetization ${\bf m}({\bf x};t) = \int d\theta {\bf e_\theta} n({\bf x},\theta;t)$. Assuming the magnetization is a Gaussian variable with variance proportional to $\rho^\xi$, as shown in Fig.~\ref{figure_fluctuations}, we obtain the equations~\cite{SI}:
\begin{gather}
\partial_t \rho = D_0 \nabla^2 \rho + \frac{v}{4}  \nabla \cdot \left( \nabla \cdot Q \right) - v \nabla \cdot {\bf m}, \label{EqRHO}\\
\partial_t {\bf m} = D_0 \nabla^2 {\bf m} + \frac{v}{8}  \begin{pmatrix}
\partial_{xx} - \partial_{yy} & 2\partial_{xy} \\
2\partial_{xy} & -\partial_{xx} + \partial_{yy}
\end{pmatrix} {\bf m}\nonumber \\
- \frac{v}{2} \left( \nabla \rho + \nabla \cdot Q \right) 
+ \gamma_0 \left[ \beta J - 1 -r \rho^\alpha - \kappa \frac{{\bf m}^2}{\rho^2} \right] {\bf m},\label{EqMAG}
\end{gather}
with the diffusion constant $D_0 = \overline{D}/4$, the self-propulsion velocity $v=\overline{D}\overline{\varepsilon}$, the ferromagnetic interaction strength $\gamma_0 = q\gamma/(q-1)$, $\kappa = (\beta J)^2 (7-3\beta J)/8 $, $\alpha = \xi - 2$, and the nematic tensor
\begin{equation}
Q = \frac{\beta J}{2\rho} \begin{pmatrix}
m_x^2 - m_y^2 & 2m_xm_y \\
2m_xm_y & - m_x^2 + m_y^2
\end{pmatrix}.
\end{equation}
Note for $r=0$, the simple mean-field theory can be recovered by neglecting the number and magnetization fluctuations. As shown in previous studies for the AIM and the APM~\cite{AIM,APM}, these mean-field equations do not predict stable phase-separated profiles and will only give the trivial homogeneous solution. We note that Eqs.~(\ref{EqRHO}-\ref{EqMAG}) allow two homogeneous solutions with $\rho = \rho_0$ corresponding to the gas phase: ${\bf m} = {\bf 0}$, and the polar liquid phase: ${\bf m}^2 = \rho_0^2 (\beta J - 1 -r \rho_0^\alpha)/\kappa$. The order-disorder transition at $\overline{\varepsilon}=0$ occurs at a density $\rho_* = [(\beta J -1)/r]^{1/\alpha}$. 

The Eqs.~\eqref{EqRHO} and~\eqref{EqMAG} are equivalent to the hydrodynamic equations derived for the VM~\cite{tonertu,solon-vm}, although the second term of the right hand side of both equations is absent due to the biased diffusion present in the model. Applying the conclusions made in Ref.~\cite{solon-vm} to our hydrodynamic equations, we are not able to conclude when a macrophase or a microphase separation is observed in the coexistence phase. Adding a zero-mean vectorial Gaussian white noise of variance $\rho^\alpha(r-\kappa{\bf m}^2/\rho^\xi)$ to the Eq.~\eqref{EqMAG} would be a possibility to scrutinize the stability of a macrophase or a microphase separation in the coexistence phase, as demonstrated for the VM in \cite{solon-vm}. Moreover, the study of the existence of reorientation transition is feasible with Eqs.~\eqref{EqRHO} and~\eqref{EqMAG}, as already done for the APM~\cite{APM}.

\medskip


{\bfseries Conclusion.} -- The nature of the flocking transition in the $q$-state ACM and the AXYM is a liquid-gas phase transition for all values of the number of states $q$, similar to the VM~\cite{VM,solon-vm}, the AIM~\cite{AIM} and the APM~\cite{Peruani11,APM}, with a coexistence phase delimiting the gas and liquid homogeneous phases for $\bar{\varepsilon}>0$. The coexistence phase shows a macrophase separation for small directions or $q$ values as in the AIM~\cite{AIM}, the APM~\cite{Peruani11,APM} and microphase separation for large $q$ values as in the VM. Longitudinally moving bands exist only when the coexistence phase is macrophase-separated, which implies that a re-orientation transition as in the APM~\cite{APM} is absent for  the ACM with large number of states and thus also for the AXYM, as it is for the VM. These results are supported by the number and magnetization fluctuations. Giant fluctuations observed for large $q$ values do not allow bulk phase separation and break large liquid domains into narrow periodic traveling bands and also restrict those bands from further coarsening, resulting in microphase separation.

Hence the discretization of the directions of motion in the VM as in the ACM will not change the characteristics of the VM flocking transition as long as the number of directions is sufficiently large. The main difference between the ACM and VM arises at zero-activity limit $\bar{\varepsilon}=0$ where the particles in the ACM can still diffuse whereas in the VM they are immobile. For a smaller number of directions, macrophase separation and a re-orientation transition occurs. The hydrodynamic description that we derived for the $q$-state ACM is compatible with the hydrodynamic description for the VM presented in Refs.~\cite{solon-vm,tonertu}, but is inconclusive regarding the stability of macrophase or microphase separation. 

Experimental realizations of various flocking models are manifold~\cite{Marchetti2013} and the small $q$ variant of the ACM (and APM) has been used to understand pattern formation observed in experiments with motility assays~\cite{Schaller2010}. For experimental systems with a large number of motility directions, the q-state ACM and the AXYM could also be a very good candidate where larger direction changes are penalized by smaller transition probabilities and a biased hopping can always be performed along the direction of motion of the particle.

When this work was finalized we became aware of a related study \cite{solon-acm} considering a version of the ACM/AXYM that differs in various important aspects from ours: in the model used in \cite{solon-acm} 1) particles live on a square lattice and hence can only move in four different directions, 2) spin flips (clock changes) can only happen to the previous or next hour, 3) the hopping rules are defined differently and are 4) projected onto the four lattice directions, which is not fully commensurate with the spin anisotropy, and 5) the hydrodynamic theory is one for XY spins in an anisotropy potential producing a term stabilizing LRO for all finite $q$-values, which is absent in our theory. For such a model an asymptotic macro-phase separation and the absence of a re-orientation transition for all $q < \infty$ is predicted in~\cite{solon-acm}. The latter is a consequence of the different hopping rules~\cite{APM}, but to numerically prove the existence or absence of an asymptotic cross-over from micro- to macro-phase separation for higher $q$-values one would have to consider much larger system sizes than those considered in~\cite{solon-acm} and by us and should be clarified in a future work.

Also, an interesting problem to investigate would be the relation between the presence of GNF in the liquid phase, the nature of the coexistence phase (micro- or macro-phase separation) and the pinned property of the spin, equivalent to a LRO or QLRO phase as a function of various control parameters.

\medskip


{\bfseries Acknowledgement.} -- This work was performed with financial support from the German Research Foundation (DFG) within the Collaborative Research Center SFB 1027. We want to thank Prof. Raja Paul for valuable discussions and for careful reading of the manuscript. SC and MM have contributed equally to the manuscript.


\newpage

\onecolumngrid

\setcounter{equation}{0}
\setcounter{figure}{0}
\renewcommand{\theequation}{S\arabic{equation}}
\renewcommand{\thefigure}{S\arabic{figure}}

\begin{center}
{\large \bfseries Supplementary Material for ``Polar flocks with discretized directions: the active clock model approaching the Vicsek model''}

\bigskip

{\normalsize Swarnajit Chatterjee,$^1$ Matthieu Mangeat,$^1$ and Heiko Rieger$^{1,2}$}

\medskip

{\normalsize \itshape $^1$Center for Biophysics \& Department for Theoretical Physics,\\ Saarland University, 66123 Saarbr{\"u}cken, Germany.\\
$^2$INM – Leibniz Institute for New Materials, Campus D2 2, 66123 Saarbrücken, Germany.}
\end{center}

\section{$q$-state ACM on Discrete Lattices}

\subsection{The Model}
\label{model}
We consider an ensemble of $N$ particles defined on periodic 2d lattices with $L_x \times L_y$ sites. The average particle density in the system is $\rho_0=N/(L_x L_y)$. Each particle endowed with a spin state (or clock angle) $\theta\in \{0,2\pi/q,4\pi/q,\cdots, 2(q-1)\pi/q\}$ can either flip to a different spin-state $\theta'$ or jump to a nearest neighbour lattice site probabilistically. The spin-state of the $k$-th particle on site $i$ is denoted $\theta_i^k$. The number of particles on site $i$ is denoted by $\rho_i$ with no restriction on its value, and the magnetization on site $i$ reads
\begin{equation}
\label{defMAG}
{\bf m}_i = \sum_{k=1}^{\rho_i} \cos \theta_i^k \ {\bf e_x} + \sum_{k=1}^{\rho_i} \sin \theta_i^k \ {\bf e_y}.
\end{equation}
The flip probabilities of the ACM are derived from a ferromagnetic Hamiltonian $H_{\rm ACM} = \sum_i H_i$ decomposed as the sum of local Hamiltonian $H_i$, taken from the standard clock model:
\begin{equation}
H_i = - \frac{J}{2\rho_i} \sum_{k=1}^{\rho_i} \sum_{l\ne k} \cos(\theta_i^k-\theta_i^l)
\end{equation}
where the prefactor $1/2\rho_i$ makes the Hamiltonian intensive and avoids the double counting of interactions and $J$ is the coupling constant between particles. When $q=2$, we recover the Hamiltonian defined for the AIM. Consider now a spin flip of a single particle on site $i$ from state $\theta$ to state $\theta'$. Without any loss of generality we can suppose that the $l^{\rm th}$ particle flips. Only the on-site energy is changed, leading to an energy difference between the new and the old state:
\begin{align}
\Delta H &= - \frac{J}{\rho_i} \sum_{k=1, k \ne l}^{\rho_i} \left[ \cos(\theta_i^k-\theta') - \cos(\theta_i^k-\theta) \right]\\
&= - \frac{J}{\rho_i} \sum_{k=1, k \ne l}^{\rho_i} \left[ \cos\theta_i^k (\cos \theta' - \cos \theta) + \sin\theta_i^k (\sin \theta' - \sin \theta) \right].
\end{align}
Defining the conserved quantity during the flip
\begin{equation}
{\bm \mu}_i = \sum_{k=1, k \ne l}^{\rho_i} \cos \theta_i^k \ {\bf e_x} + \sum_{k=1, k \ne l}^{\rho_i} \sin \theta_i^k \ {\bf e_y},
\end{equation}
we get the energy difference:
\begin{equation}
\Delta H = - \frac{J}{\rho_i} {\bm \mu}_i \cdot ({\bf e_{\theta'}} - {\bf e_\theta}).
\end{equation}
From the Eq.~\eqref{defMAG}, ${\bm \mu}_i$ is linked to the magnetization on the site $i$ before the flip ${\bf m}_i$ and after the flip ${\bf m}_i'$ with the relations ${\bm \mu}_i = {\bf m}_i - {\bf e_\theta} = {\bf m}_i' - {\bf e_{\theta'}}$, which lead to the energy difference:
\begin{equation}
\Delta H = - \frac{J}{\rho_i} \left[ {\bf m}_i \cdot ({\bf e_{\theta'}} - {\bf e_\theta}) + 1 - \cos(\theta'-\theta) \right],
\end{equation}
used in the main text. The energy difference can also be written as
\begin{equation}
\Delta H = - \frac{J}{2\rho_i} \left( {\bf m}_i + {\bf m}_i'\right) \cdot ({\bf e_{\theta'}} - {\bf e_\theta}),
\end{equation}
with both the magnetizations before and after the flip.

In analogy to the AIM and the APM, the transition rate is chosen to verify the detailed balance:
\begin{equation}
\label{defWflip}
W_{\rm flip}(\theta,\theta') = \gamma \exp(-\beta \Delta H) =  \gamma \exp\left\{\frac{\beta J}{\rho_i} \left[ {\bf m}_i \cdot ({\bf e_{\theta'}} - {\bf e_\theta}) + 1 - \cos(\theta'-\theta) \right]\right\}.
\end{equation}
Moreover, each particle performs a biased diffusion on the lattice depending on the particle state $\theta$: the hopping rate is $W_{\rm hop} = D(1 + \varepsilon)$ in the direction $\theta$ and $W_{\rm hop} = D[1 -\varepsilon/(q - 1)]$, otherwise.

We perform the numerical simulations with a Monte Carlo algorithm similar to the one used to analyze the active Potts model (APM)~\cite{apm} using the flip rate derived in Eq.~(\ref{defWflip}). Here we study the $q=4$-state ACM and the $q=6$-state ACM on a 2d square lattice and 2d triangular lattice, respectively. The time is discretized in small time units $\Delta t=[qD+\exp(2 \beta J)]^{-1}$ where the time increment is defined as $\Delta t /N$, $N$ being the total number of particles. At each $\Delta t /N$, a randomly chosen particle either flips its state from $\theta$ to $\theta'$ with probability $W_{\rm flip}(\theta,\theta')\Delta t$ or hops to a nearest neighbor with probability $W_{\rm hop}\Delta t$.

\subsection{Numerical Results}
\label{result}
Now, we will present numerical results from our simulations of the 4-state ACM. Simulations are performed on a $100 \times 100$ square lattice using three control parameters: the temperature $T=\beta^{-1}$, the average particle density $\rho_0$, and the bias $\varepsilon$.
\begin{figure}[H]
\centering
\includegraphics[width=0.55\columnwidth]{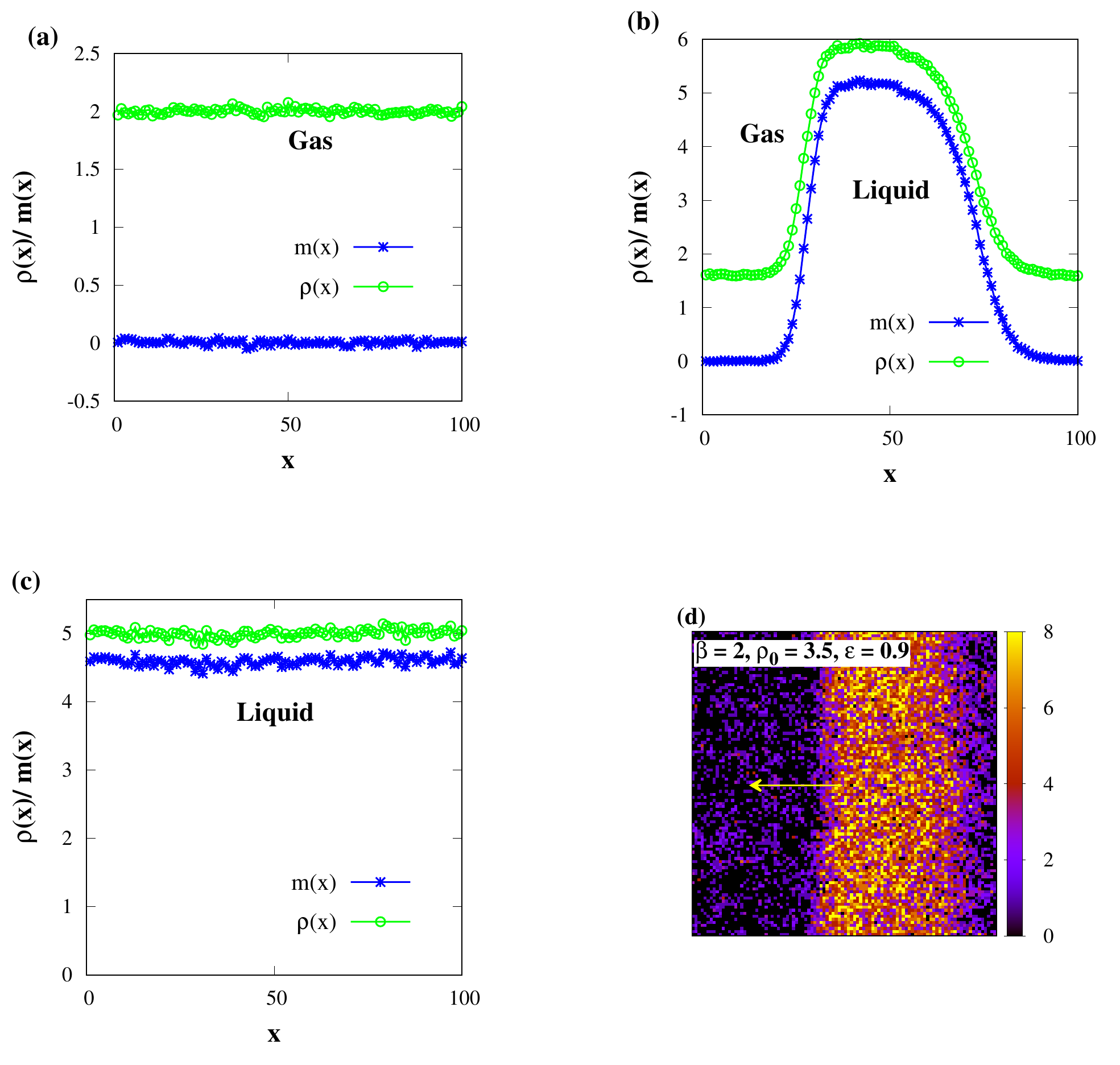}
\caption{(color online) Three phases of the 4-state ACM for $\varepsilon=0.9$, (a) disordered gas for $\beta=1.2$, $\rho_0=2$, (b) liquid-gas co-existence for $\beta=2$, $\rho_0=3.5$, and (c) polar liquid for $\beta=2.3$, $\rho_0=5$. (d) Density field snapshot corresponding to (b).}
\label{fig1}
\end{figure}

The three typical phases of the ACM are shown in Fig.~\ref{fig1} for $\varepsilon=0.9$. A disordered gaseous phase at high temperature and low density ($\beta=1.2$, $\rho_0=2$) in Fig.~\ref{fig1}(a) are followed by a liquid-gas co-existence phase in Fig.~\ref{fig1}(b) for intermediate temperature and density ($\beta=2$, $\rho_0=3.5$) and a polar liquid phase in Fig.~\ref{fig1}(c) at low temperature and high density ($\beta=2.3$, $\rho_0=5$). In Fig.~\ref{fig1}(d), we represent the corresponding snapshot of Fig.~\ref{fig1}(b) where a fully phase-separated polar liquid band is shown traveling transversely on a gaseous background. All the profiles presented in Fig.~\ref{fig1}(a--c) are averaged over space and time and the two homogeneous phases, gas and liquid, are defined respectively by the average magnetization where for a gas phase $\langle m \rangle \approx 0$ and for the liquid  phase, $\langle m \rangle \approx m_0 \neq 0$.

\begin{figure}[H]
\centering
\includegraphics[width=0.7\columnwidth]{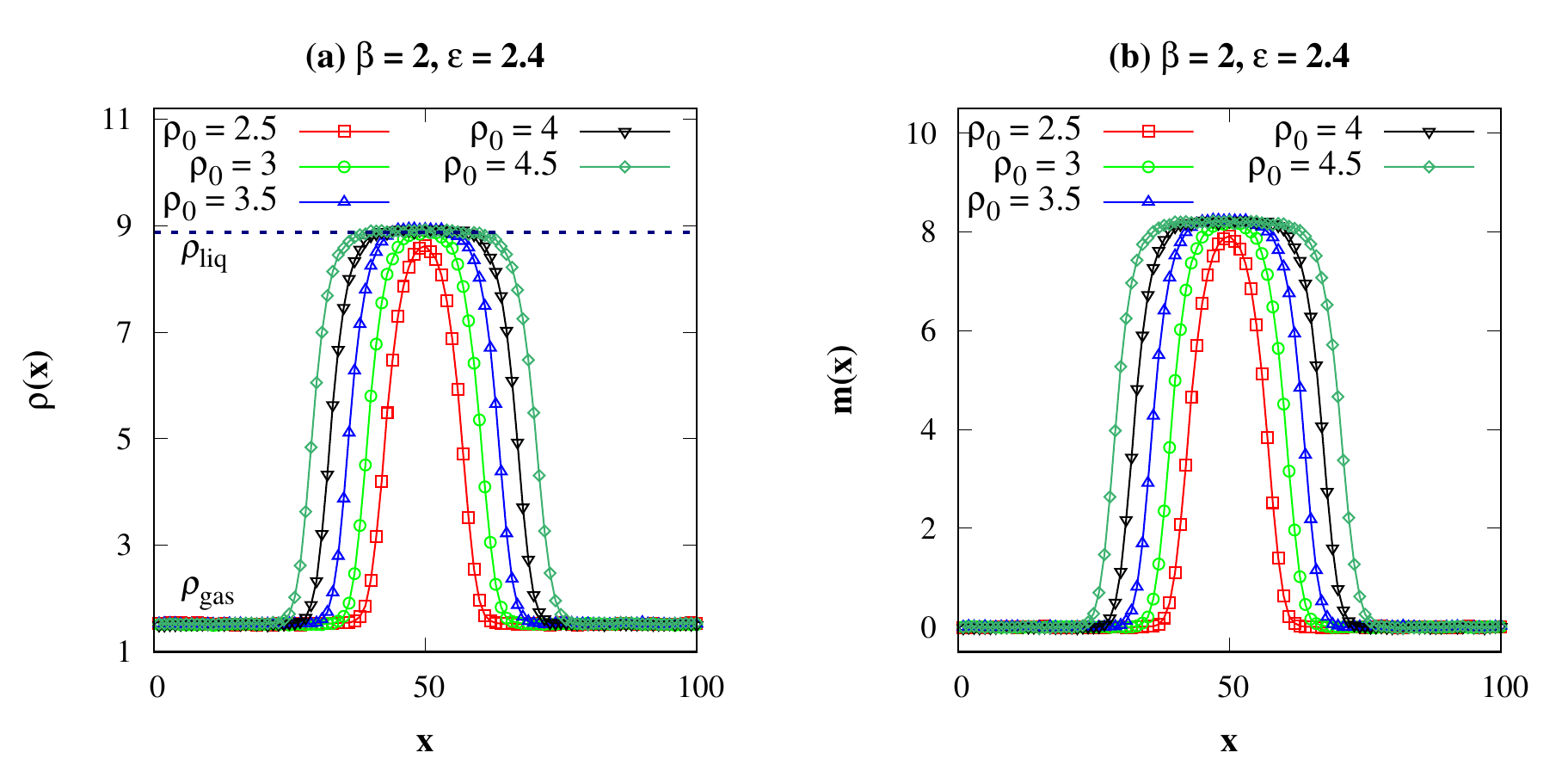}
\caption{(color online) Segregated (a) density and (b) magnetization profiles with increasing initial $\rho_0$ are shown for $\beta=2$ and $\varepsilon=2.4$.} 
\label{fig2}
\end{figure}

Phase-separated density and magnetization profiles (averaged along the $y$-axis and over time) of the liquid-gas coexistence phase are shown in Fig.~\ref{fig2}(a) and Fig.~\ref{fig2}(b) respectively, for $\beta=2, \varepsilon=2.4$ and several $\rho_0$. The width of the polar liquid band increases with the average density $\rho_0$ without affecting the densities of the liquid $\rho_{\rm liq}(T,\varepsilon)$ and the gaseous $\rho_{\rm gas}(T,\varepsilon)$ phases. A single internal state $(\theta=\pi/2)$ dominates each of the band, then all these bands are longitudinal in nature.

\begin{figure}[H]
\centering
\includegraphics[width=0.6\columnwidth]{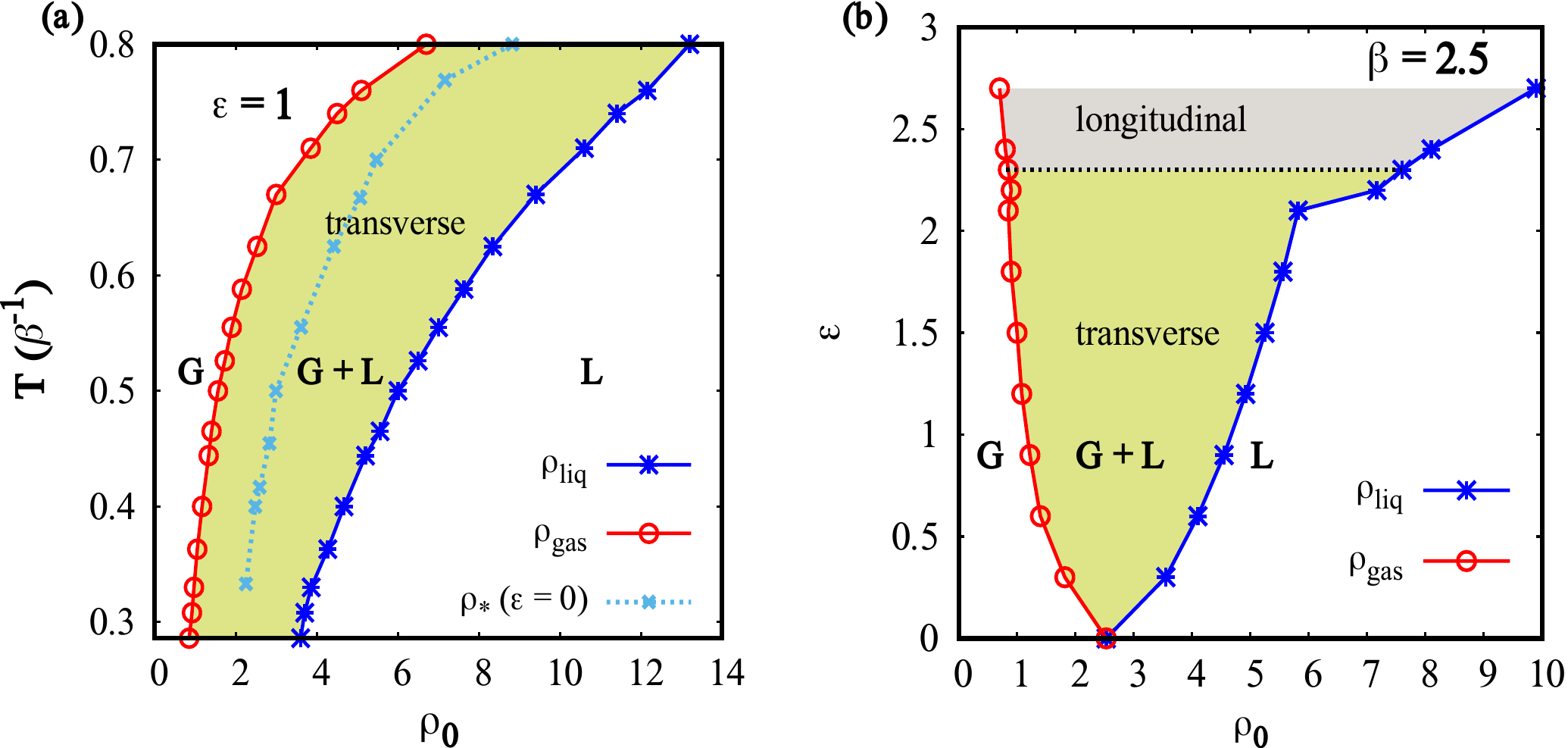}
\caption{(color online) Phase diagrams of the 4-state ACM. (a) Temperature-density ($T$-$\rho_0$) phase diagram for $\varepsilon=1$. The dotted line indicates the transition density $\rho^*$ for $\varepsilon=0$. (b) Velocity-density ($\varepsilon$-$\rho_0$) diagram for $\beta=2.5$ where the black dotted line indicates the reorientation transition line from transverse to longitudinal particle motion.} 
\label{fig3}
\end{figure}

\begin{figure}[H]
\centering
\includegraphics[width=0.6\columnwidth]{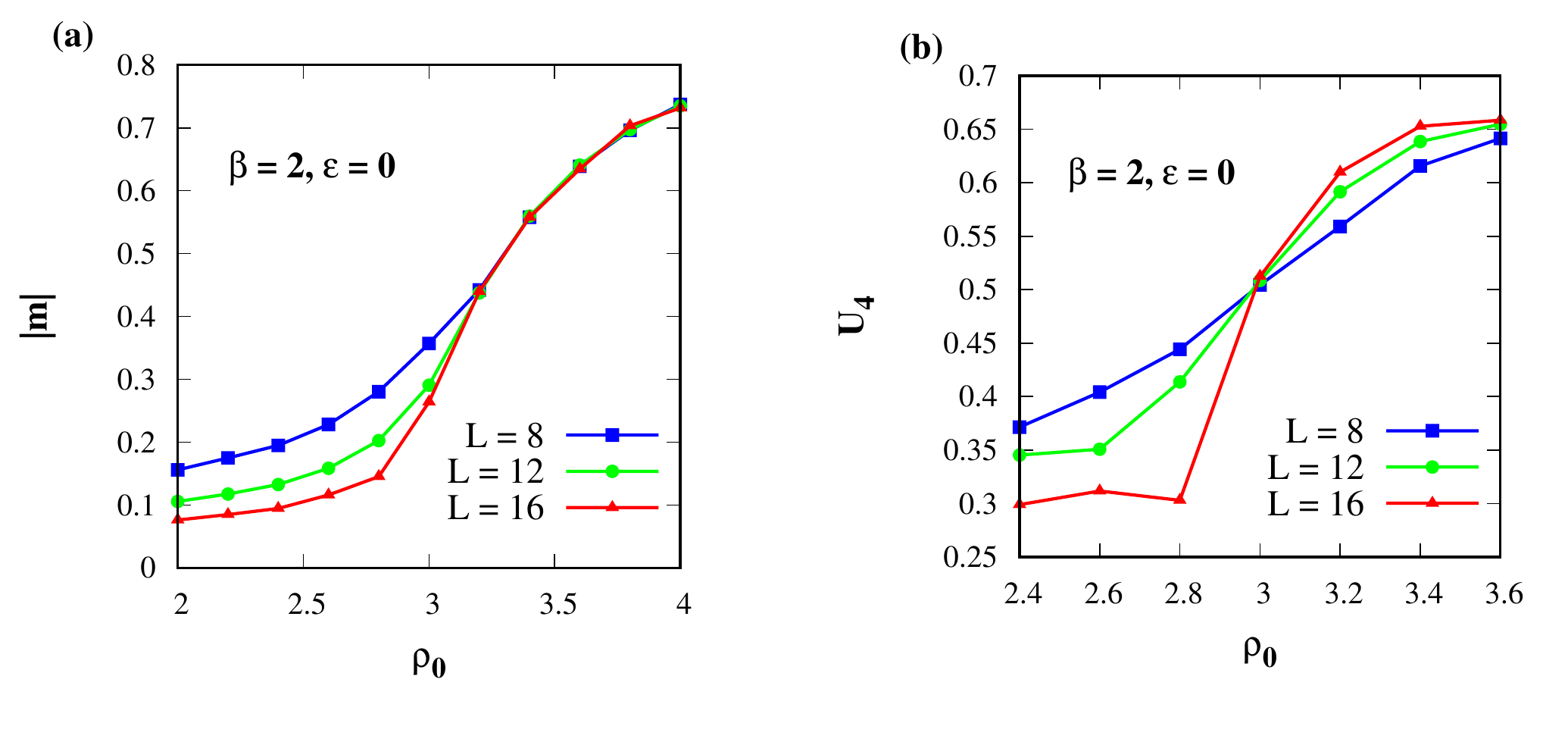}
\caption{(color online) Liquid-gas phase transition in 4-state ACM for $\beta=2$ and $\varepsilon=0$. (a)~Magnetization $|{\bf m}|$ versus $\rho_0$ for different lattice sizes $L = 8$, $L=12$, and $L=16$. (b)~Binder cumulant $U_4$ versus $\rho_0$ for different lattice sizes $L$. The critical density $\rho_* (\beta=2) = 2.99 \pm 0.01$ is extracted from the intersection of these curves.} 
\label{fig4}
\end{figure}

In Fig.~\ref{fig3}(a) we show the phase diagram of the 4-state ACM in the $(T,\rho_0)$ plane for $\varepsilon=1$. The binodals $\rho_{\rm gas}$ and $\rho_{\rm liq}$, which are are computed from the time averaged phase separated density profiles shown in Fig.~\ref{fig2}, segregate the gaseous ($G$), gas-liquid co-existence ($G+L$), and liquid ($L$) phases. The dashed line inside the co-existence region represents the critical densities $\rho_* (\beta)$ where the liquid-gas transition occurs at $\varepsilon=0$ (see Fig.~\ref{fig4}). The ($\varepsilon$,$\rho_0$) phase diagram for a fixed temperature $\beta=2.5$ is shown in Fig.~\ref{fig3}(b). $\rho_{\rm gas}$ and $\rho_{\rm liq}$ merges at $\rho_*(\beta=2.5,\varepsilon=0) \simeq 2.5$ for $\varepsilon=0$, which is the critical point. We have already demonstrated in Fig.~\ref{fig1} and Fig.~\ref{fig2} that the 4-state ACM exhibits the reorientation transition of the co-existence phase and it is depicted in Fig.~\ref{fig3}(b) through two different color shades. In the $(T,\rho_0)$ phase diagram we do not observe the longitudinal phase as the diagram is obtained for small particle velocity, $\varepsilon=1$ for which the system manifests only transverse band motion. In the ($\varepsilon$,$\rho_0$) phase diagram the transition approximately happens at $\varepsilon \simeq 2.3$ (represented by black dotted line).

The $\varepsilon=0$ limit of the ACM is the purely diffusive version of the model where a continuous phase transition is observed from a low-density homogeneous phase to a high-density ordered phase without the gas-liquid coexistence phase as presented in Fig.~\ref{fig4} for $\beta=2$. Such second order transition was also observed in the AIM~\cite{aim} but in the APM~\cite{apm}, the transition reported was first order in nature. In Fig.~\ref{fig4}(a), the magnetization is plotted against $\rho_0$ and we observe a smooth, continuous transition from a high magnetized liquid state at larger $\rho_0$ to a gaseous state at smaller $\rho_0$. The critical density of this transition, $\rho^*$ is calculated from the Binder cumulant $U_4 = 1-\langle |{\bf m}|^4 \rangle/3 \langle |{\bf m}|^2 \rangle^2$ versus $\rho_0$ shown in Fig.~\ref{fig4}(b) and from the intersection of the $U_4$ curves for different $L$, we quantified the critical density $\rho_* (\beta=2) = 2.99 \pm 0.01$.

\begin{figure}[H]
\centering
\includegraphics[width=0.58\columnwidth]{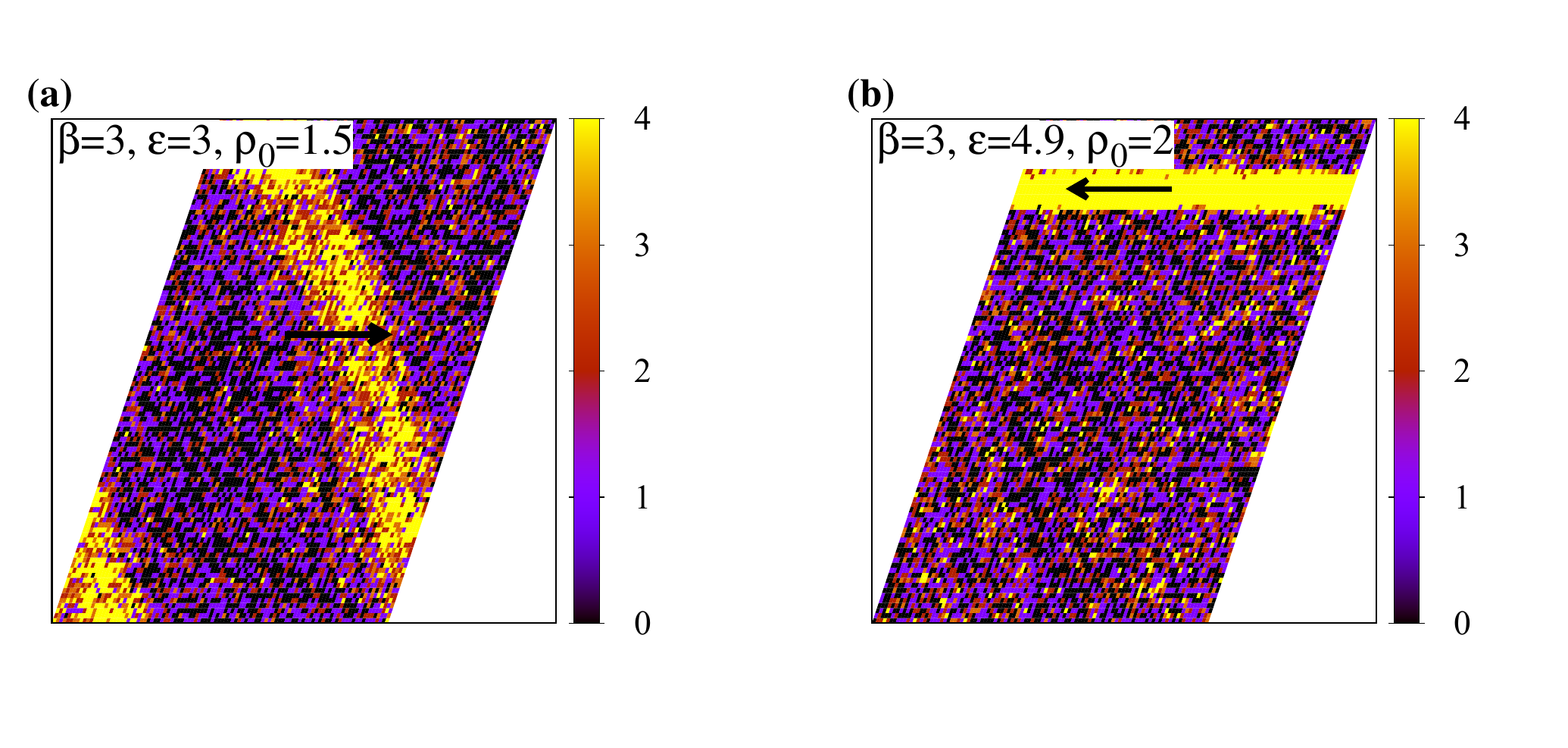}
\caption{(color online) Steady-state density snapshots of the 6-state ACM on a triangular lattice of dimension $100 \times 100$ showing (a) transverse band motion for $\varepsilon=3$, $\rho_0=1.5$ and (b) longitudinal band motion for $\varepsilon=4.9$, $\rho_0=2$. $\beta=3$. Colorbar represents site occupation.} 
\label{fig5}
\end{figure}

Additionally, we present two snapshots of the 6-state ACM on a triangular lattice in Fig.~\ref{fig5}(a--b) as a function of $\varepsilon$ confirming the band to lane reorientation transition also for $q=6$. In Fig.~\ref{fig5}(a) we show the transverse motion of the polar liquid band for $\varepsilon=3$, which is constituted by particles having internal state $\theta=0$ whereas longitudinal lane formation along the predominant direction of the particles with $\theta=\pi$ is observed for $\varepsilon=4.9$. It was shown in the context of APM~\cite{apm} that this reorientation transition which was not present in other known flocking models, is not an artefact of any algorithmic implementation and our investigation of the $q$-state ACM on discrete lattices has validated that argument further.

\section{Hydrodynamic description}

Now, we will present the derivation of the hydrodynamic equations~(3)-(5) presented in the main text. We define $n({\bf x},\theta;t)$ as the probability density for a particle to be at the position ${\bf x}$ and in the spin-state $\theta$ at the time $t$. The particle density reads
\begin{equation}
\rho({\bf x},t) = \int_0^{2\pi} d\theta n({\bf x},\theta;t) = \Delta \theta \sum_\theta n({\bf x},\theta;t)
\end{equation}
for the AXYM and the $q$-state ACM with $\Delta \theta = 2 \pi /q$, respectively. Similarly, the magnetization is defined by
\begin{equation}
{\bf m}({\bf x},t) = \int_0^{2\pi} d\theta {\bf e_\theta} n({\bf x},\theta;t) = \Delta \theta \sum_\theta {\bf e_\theta} n({\bf x},\theta;t),
\end{equation}
with ${\bf e_\theta} = (\cos \theta, \sin \theta)$. Note that $\rho({\bf x}_i,t) \equiv \rho_i(t)$ and ${\bf m}({\bf x}_i,t) \equiv {\bf m}_i(t)$ represents the particle number and the magnetization in the neighborhood ${\cal N}_i$, respectively. Finally, we also define the nematic tensor as 
\begin{equation}
Q({\bf x},t) = \int_0^{2\pi} d\theta \begin{pmatrix}
\cos 2\theta & \sin 2\theta \\
\sin 2\theta & - \cos 2\theta
\end{pmatrix} n({\bf x},\theta;t) = \Delta \theta \sum_\theta \begin{pmatrix}
\cos 2\theta & \sin 2\theta \\
\sin 2\theta & - \cos 2\theta
\end{pmatrix} n({\bf x},\theta;t) \equiv
\begin{pmatrix}
Q_1({\bf x},t) & Q_2({\bf x},t) \\
Q_2({\bf x},t) & - Q_1({\bf x},t)
\end{pmatrix},
\end{equation}
and we suppose that higher harmonic terms are zero.

We derive the master equation corresponding to the microscopic process with hopping rates $W_{\rm hop}(\theta,\phi)$ and flipping rates $W_{\rm flip}(\theta,\theta')$, where $\theta$, $\phi$ and $\theta'$ are the state angle of the particle, the hopping angle and the angle after the flip, respectively. It writes then as
\begin{gather}
n({\bf x},\theta;t+dt) = n({\bf x},\theta;t) \left[ 1 - dt \sum_\phi W_{\rm hop}(\theta,\phi) - dt \sum_{\theta' \ne \theta} \frac{W_{\rm flip}(\theta,\theta')}{q-1} \right] \nonumber \\
+ \left[ \sum_\phi n({\bf x}-{\bf e_\phi},\theta;t)W_{\rm hop}(\theta,\phi) + \sum_{\theta' \ne \theta} n({\bf x},\theta';t) \frac{W_{\rm flip}(\theta',\theta)}{q-1}\right] dt.
\end{gather}
Taking the limit $dt \to 0$, we get
\begin{equation}
\frac{\partial n}{\partial t}({\bf x}) = \sum_\phi \left[n({\bf x}-{\bf e_\phi},\theta) - n({\bf x},\theta) \right]W_{\rm hop}(\theta,\phi) + \frac{1}{q-1}\sum_{\theta' \ne \theta} \left[ n({\bf x},\theta') W_{\rm flip}(\theta',\theta) - n({\bf x},\theta)  W_{\rm flip}(\theta,\theta') \right]
\equiv I_{\rm hop} + I_{\rm flip}.
\end{equation}
We can obviously add the term $\theta=\theta'$ in $I_{\rm flip}$. First, we calculate the expression of $I_{\rm hop}$. Using the definition of $W_{\rm flip}$, we obtain
\begin{equation}
I_{\rm hop} = \frac{\overline{D}(1-\overline{\varepsilon})}{q} \sum_\phi \left[n({\bf x}-{\bf e_\phi},\theta) - n({\bf x},\theta) \right] + \overline{D}\overline{\varepsilon} \left[n({\bf x}-{\bf e_\theta},\theta) - n({\bf x},\theta) \right].
\end{equation}
In the hydrodynamic limit, we show that
\begin{equation}
n({\bf x}-{\bf e_\phi},\theta) - n({\bf x},\theta) = - {\bf e_\phi} \cdot \nabla n({\bf x},\theta) + \frac{1}{2} \left( {\bf e_\phi} \cdot \nabla \right)^2 n({\bf x},\theta) + \cdots,
\end{equation}
and we deduce then
\begin{equation}
\sum_\phi \left[n({\bf x}-{\bf e_\phi},\theta) - n({\bf x},\theta) \right] = \frac{q}{4} \nabla^2 n({\bf x},\theta).
\end{equation}
The hopping term becomes
\begin{equation}
I_{\rm hop} = \frac{\overline{D}(1-\overline{\varepsilon})}{4} \nabla^2 n({\bf x},\theta) + \frac{\overline{D}\overline{\varepsilon}}{2} \left( {\bf e_\theta} \cdot \nabla \right)^2 n({\bf x},\theta) - \overline{D}\overline{\varepsilon} {\bf e_\theta} \cdot \nabla n({\bf x},\theta).
\end{equation}
Merging the diffusive terms together, we obtain
\begin{equation}
I_{\rm hop} = \frac{\overline{D}}{4} \nabla^2 n({\bf x},\theta) + \frac{\overline{D}\overline{\varepsilon}}{4} \nabla \cdot \begin{pmatrix}
\cos 2\theta & \sin 2\theta \\
\sin 2\theta & - \cos 2\theta
\end{pmatrix} \nabla n({\bf x},\theta) - \overline{D}\overline{\varepsilon} {\bf e_\theta} \cdot \nabla n({\bf x},\theta).
\end{equation}

Now, we calculate the flipping term $I_{\rm flip}$. We only keep the first-order terms in the $|{\bf m}_i| \ll \rho_i$ expansion. At the leading order, we may suppose
\begin{equation}
W_{\rm flip}(\theta,\theta') \simeq \gamma \exp\left[ \frac{\beta J}{\rho} {\bf m} \cdot \left( {\bf e_\theta} - {\bf e_{\theta'}} \right) \right]
\end{equation}
and the Taylor expansion gives
\begin{equation}
W_{\rm flip}(\theta,\theta') \simeq \gamma \left[ 1 + \frac{\beta J}{\rho} {\bf m} \cdot \left( {\bf e_\theta} - {\bf e_{\theta'}} \right) + \frac{1}{2} \left(\frac{\beta J}{\rho}\right)^2 \left[{\bf m} \cdot \left( {\bf e_\theta} - {\bf e_{\theta'}} \right) \right]^2 + \frac{1}{6} \left(\frac{\beta J}{\rho}\right)^3 \left[{\bf m} \cdot \left( {\bf e_\theta} - {\bf e_{\theta'}} \right) \right]^3 \right].
\end{equation}
Then, the flipping term writes
\begin{gather}
I_{\rm flip} = \frac{\gamma}{q-1} \sum_{\theta'} \left\{ \left[ n({\bf x},\theta') - n({\bf x},\theta) \right] - \frac{\beta J}{\rho} \left[{\bf m} \cdot \left( {\bf e_{\theta'}} - {\bf e_{\theta}} \right) \right] \left[ n({\bf x},\theta') + n({\bf x},\theta) \right] \right. \nonumber \\
\left.+ \frac{1}{2} \left(\frac{\beta J}{\rho} \right)^2 \left[{\bf m} \cdot \left( {\bf e_{\theta'}} - {\bf e_{\theta}} \right) \right]^2 \left[ n({\bf x},\theta') - n({\bf x},\theta) \right] - \frac{1}{6} \left(\frac{\beta J}{\rho} \right)^3 \left[{\bf m} \cdot \left( {\bf e_{\theta'}} - {\bf e_{\theta}} \right) \right]^3 \left[ n({\bf x},\theta') + n({\bf x},\theta) \right] \right\}.\label{eqIFLIP}
\end{gather}
The first term in Eq.~\eqref{eqIFLIP} writes
\begin{equation}
\sum_{\theta'} \left[ n({\bf x},\theta') - n({\bf x},\theta) \right] = \frac{q}{2\pi} \left[\rho - 2 \pi n \right],
\end{equation}
where we have simplified the notations: $\rho \equiv \rho({\bf x})$ and $n \equiv n({\bf x},\theta)$. The second term in Eq.~\eqref{eqIFLIP} writes
\begin{equation}
\sum_{\theta'} \left[{\bf m} \cdot \left( {\bf e_{\theta'}} - {\bf e_{\theta}} \right) \right] \left[ n({\bf x},\theta') + n({\bf x},\theta) \right] = \frac{q}{2\pi} \left[{\bf m}^2 - \left( \rho + 2 \pi n \right) ( {\bf m} \cdot {\bf e_\theta} ) \right],
\end{equation}
where ${\bf m} \equiv {\bf m}(\bf x)$. The third term in Eq.~\eqref{eqIFLIP} writes
\begin{equation}
\sum_{\theta'} \left[{\bf m} \cdot \left( {\bf e_{\theta'}} - {\bf e_{\theta}} \right) \right]^2 \left[ n({\bf x},\theta') - n({\bf x},\theta) \right] = \frac{q}{2\pi} \left[ \frac{1}{2} {\bf m} \cdot Q {\bf m} - 2 {\bf m}^2 ({\bf m} \cdot {\bf e_\theta}) + \frac{1}{2} \left( \rho - 2 \pi n \right) \left( {\bf m}^2 + 2( {\bf m} \cdot {\bf e_\theta} )^2 \right) \right],
\end{equation}
where $Q \equiv Q(\bf x)$. The fourth term in Eq.~\eqref{eqIFLIP} writes
\begin{equation}
\sum_{\theta'} \left[{\bf m} \cdot \left( {\bf e_{\theta'}} - {\bf e_{\theta}} \right) \right]^3 \left[ n({\bf x},\theta') + n({\bf x},\theta) \right] = -\frac{3q}{4\pi} \left[ {\bf m} \cdot Q {\bf m} + \left( \rho + 2 \pi n \right) \left( {\bf m}^2 + \frac{2}{3}( {\bf m} \cdot {\bf e_\theta} )^2 \right) \right] ( {\bf m} \cdot {\bf e_\theta} ).
\end{equation}
Merging all these terms together, the Eq.~\eqref{eqIFLIP} becomes
\begin{gather}
I_{\rm flip} = \frac{q\gamma}{2\pi(q-1)} \left\{ \left[ \rho - 2 \pi n  \right] - \frac{\beta J}{\rho} \left[{\bf m}^2 - \left( \rho + 2 \pi n \right) ( {\bf m} \cdot {\bf e_\theta} ) \right] \right. \nonumber \\
+ \frac{1}{2} \left(\frac{\beta J}{\rho} \right)^2 \left[ \frac{1}{2} {\bf m} \cdot Q {\bf m} - 2 {\bf m}^2 ({\bf m} \cdot {\bf e_\theta}) + \frac{1}{2} \left( \rho - 2 \pi n \right) \left( {\bf m}^2 + 2( {\bf m} \cdot {\bf e_\theta} )^2 \right) \right] \nonumber \\
\left.+ \frac{1}{4} \left(\frac{\beta J}{\rho} \right)^3 \left[ {\bf m} \cdot Q {\bf m} + \left( \rho + 2 \pi n \right) \left( {\bf m}^2 + \frac{2}{3}( {\bf m} \cdot {\bf e_\theta} )^2 \right) \right] ( {\bf m} \cdot {\bf e_\theta} ) \right\},\label{Iflip}
\end{gather}
and the hydrodynamic equation writes
\begin{gather}
\label{eqME}
\frac{\partial n}{\partial t} = \frac{\overline{D}}{4} \nabla^2 n + \frac{\overline{D}\overline{\varepsilon}}{4} \nabla \cdot \begin{pmatrix}
\cos 2\theta & \sin 2\theta \\
\sin 2\theta & - \cos 2\theta
\end{pmatrix} \nabla n
- \overline{D}\overline{\varepsilon} {\bf e_\theta} \cdot \nabla n \nonumber \\
+ \frac{q\gamma}{2\pi(q-1)} \left\{ \left[ \rho - 2 \pi n  \right] - \frac{\beta J}{\rho} \left[{\bf m}^2 - \left( \rho + 2 \pi n \right) ( {\bf m} \cdot {\bf e_\theta} ) \right] \right. \nonumber \\
+ \frac{1}{2} \left(\frac{\beta J}{\rho} \right)^2 \left[ \frac{1}{2} {\bf m} \cdot Q {\bf m} - 2 {\bf m}^2 ({\bf m} \cdot {\bf e_\theta}) + \frac{1}{2} \left( \rho - 2 \pi n \right) \left( {\bf m}^2 + 2( {\bf m} \cdot {\bf e_\theta} )^2 \right) \right] \nonumber \\
\left.+ \frac{1}{4} \left(\frac{\beta J}{\rho} \right)^3 \left[ {\bf m} \cdot Q {\bf m} + \left( \rho + 2 \pi n \right) \left( {\bf m}^2 + \frac{2}{3}( {\bf m} \cdot {\bf e_\theta} )^2 \right) \right] ( {\bf m} \cdot {\bf e_\theta} ) \right\}.
\end{gather}
This equation for $n({\bf x},\theta)$ depends on the integrated quantities $\rho({\bf x})$,  ${\bf m}({\bf x})$ and  $Q({\bf x})$. To have closed equations, we derive now the equations for these integrated functions.

The density $\rho({\bf x})$ fulfills the equation
\begin{equation}
\partial_t \rho = \frac{\overline{D}}{4} \nabla^2 \rho + \frac{\overline{D}\overline{\varepsilon}}{4}  \nabla \cdot \left( \nabla \cdot Q \right) - \overline{D}\overline{\varepsilon} \nabla \cdot {\bf m}
\end{equation}
since $\sum_\theta I_{\rm flip} =0$ (statement also verified from Eq.~\eqref{Iflip}).

We now derive the equation for the magnetization ${\bf m}({\bf x})$. First, we obtain
\begin{equation}
\Delta \theta \sum_\theta {\bf e_\theta} I_{\rm hop} = \frac{\overline{D}}{4} \nabla^2 {\bf m} + \frac{\overline{D}\overline{\varepsilon}}{8}  \begin{pmatrix}
\partial_{xx} - \partial_{yy} & 2\partial_{xy} \\
2\partial_{xy} & -\partial_{xx} + \partial_{yy}
\end{pmatrix} {\bf m}
- \frac{\overline{D}\overline{\varepsilon}}{2} \left( \nabla \rho + \nabla \cdot Q \right).
\end{equation}
Then, to calculate $\Delta \theta \sum_\theta {\bf e_\theta} I_{\rm flip}$, we need the expressions of
\begin{gather}
\Delta \theta \sum_\theta {\bf e_\theta} ({\bf m} \cdot {\bf e_\theta}) =\pi {\bf m}, \qquad \Delta \theta \sum_\theta n {\bf e_\theta} ({\bf m} \cdot {\bf e_\theta}) = \frac{1}{2}(\rho + Q ) {\bf m}, \\
\Delta \theta \sum_\theta {\bf e_\theta} ({\bf m} \cdot {\bf e_\theta})^2 = 0, \qquad \Delta \theta \sum_\theta n {\bf e_\theta} ({\bf m} \cdot {\bf e_\theta})^2 = \frac{3}{4} {\bf m}^2 {\bf m}, \\
\Delta \theta \sum_\theta {\bf e_\theta} ({\bf m} \cdot {\bf e_\theta})^3 =  \frac{3\pi}{4} {\bf m}^2 {\bf m}, \qquad \Delta \theta \sum_\theta n {\bf e_\theta} ({\bf m} \cdot {\bf e_\theta})^3 = \frac{3}{8} \rho {\bf m}^2 {\bf m}.
\end{gather}
The first term in Eq.~\eqref{Iflip} gives
\begin{equation}
\Delta \theta \sum_\theta {\bf e_\theta} \left[ \rho - 2 \pi n  \right] = - 2 \pi {\bf m},
\end{equation}
the second term in Eq.~\eqref{Iflip} gives
\begin{equation}
\Delta \theta \sum_\theta {\bf e_\theta} \left[{\bf m}^2 - \left( \rho + 2 \pi n \right) ( {\bf m} \cdot {\bf e_\theta} ) \right] = - 2 \pi \rho {\bf m} - \pi Q {\bf m},
\end{equation}
the third term in Eq.~\eqref{Iflip} gives
\begin{equation}
\Delta \theta \sum_\theta {\bf e_\theta} \left[ \frac{1}{2} {\bf m} \cdot Q {\bf m} - 2 {\bf m}^2 ({\bf m} \cdot {\bf e_\theta}) + \frac{1}{2} \left( \rho - 2 \pi n \right) \left( {\bf m}^2 + 2( {\bf m} \cdot {\bf e_\theta} )^2 \right) \right] = - \frac{9 \pi}{2} {\bf m}^2 {\bf m},
\end{equation}
and the forth term in Eq.~\eqref{Iflip} gives
\begin{equation}
\Delta \theta \sum_\theta {\bf e_\theta} \left[ {\bf m} \cdot Q {\bf m} + \left( \rho + 2 \pi n \right) \left( {\bf m}^2 + \frac{2}{3}( {\bf m} \cdot {\bf e_\theta} )^2 \right) \right] ( {\bf m} \cdot {\bf e_\theta} ) = 3\pi \rho {\bf m}^2 {\bf m}.
\end{equation}
We obtain then
\begin{gather}
\Delta \theta \sum_\theta {\bf e_\theta} I_{\rm flip} = \frac{q\gamma}{q-1} \left[ (\beta J - 1) - \frac{3}{2} \left(\frac{\beta J}{2\rho} \right)^2 (3-\beta J) {\bf m}^2 + \frac{\beta J}{2\rho}  Q \right] {\bf m},
\end{gather}
and the magnetization ${\bf m}({\bf x})$ fulfills the equation
\begin{gather}
\partial_t {\bf m} = \frac{\overline{D}}{4} \nabla^2 {\bf m} + \frac{\overline{D}\overline{\varepsilon}}{8}  \begin{pmatrix}
\partial_{xx} - \partial_{yy} & 2\partial_{xy} \\
2\partial_{xy} & -\partial_{xx} + \partial_{yy}
\end{pmatrix} {\bf m}
- \frac{\overline{D}\overline{\varepsilon}}{2} \left( \nabla \rho + \nabla \cdot Q \right) \nonumber \\
+ \frac{q\gamma}{q-1} \left[ (\beta J - 1) - \frac{3}{2} \left(\frac{\beta J}{2\rho} \right)^2 (3-\beta J) {\bf m}^2 + \frac{\beta J}{2\rho}  Q \right] {\bf m}.
\end{gather}
The expression of the nematic tensor $Q({\bf x})$ is obtained by neglecting the diffusion and drift terms. From Eq.~\eqref{eqME}, the nematic tensor fulfills then
\begin{equation}
\dot Q = \frac{q \gamma}{2 \pi (q-1)} \left[ -2 \pi Q + \frac{\beta J}{\rho} \Delta \theta \sum_\theta \begin{pmatrix}
\cos 2\theta & \sin 2\theta \\
\sin 2\theta & - \cos 2\theta
\end{pmatrix} (\rho+2\pi n) ({\bf m} \cdot {\bf e_\theta}) \right]. 
\end{equation}
Assuming $\dot Q = 0$, we obtain
\begin{equation}
Q = \frac{\beta J}{2\rho} \begin{pmatrix}
m_x^2 - m_y^2 & 2m_xm_y \\
2m_xm_y & - m_x^2 + m_y^2
\end{pmatrix}. 
\end{equation}

We can deduce then
\begin{equation}
Q{\bf m} = \frac{\beta J}{2\rho} {\bf m}^2 {\bf m},
\end{equation}
and the equation for the magnetization becomes
\begin{gather}
\partial_t {\bf m} = \frac{\overline{D}}{4} \nabla^2 {\bf m} + \frac{\overline{D}\overline{\varepsilon}}{8}  \begin{pmatrix}
\partial_{xx} - \partial_{yy} & 2\partial_{xy} \\
2\partial_{xy} & -\partial_{xx} + \partial_{yy}
\end{pmatrix} {\bf m}
- \frac{\overline{D}\overline{\varepsilon}}{2} \left( \nabla \rho + \nabla \cdot Q \right) \nonumber \\
+ \frac{q\gamma}{q-1} \left[ (\beta J - 1) - \frac{1}{2} \left(\frac{\beta J}{2\rho} \right)^2 (7-3\beta J) {\bf m}^2 \right] {\bf m}.
\end{gather}

The equations obtained for the density and the magnetization are not yet averaged over stochastic realizations. We consider that the magnetization follows a Gaussian distribution such that
\begin{equation}
P({\bf m}) = \frac{1}{2 \pi \sigma^2} \exp \left[ - \frac{({\bf m}-\bm{\mu})^2}{2\sigma^2}\right],
\end{equation}
where $\langle {\bf m} \rangle = \bm{\mu}$, $\langle {\bf m}^2 \rangle = \bm{\mu}^2 + 2\sigma^2$. We obtain then $\langle {\bf m}^2 {\bf m} \rangle = \bm{\mu}^2 \bm{\mu} + 4\sigma^2 \bm{\mu}$, and
\begin{equation}
Q = \frac{\beta J}{2\rho} \begin{pmatrix}
\mu_x^2 - \mu_y^2 & 2\mu_x\mu_y \\
2\mu_x\mu_y & - \mu_x^2 + \mu_y^2
\end{pmatrix}. 
\end{equation}

We have shown in the main text that $\sigma \propto \rho^\xi$ (with $1 \leqslant \xi \leqslant 1.67$), which gives the hydrodynamic equations:
\begin{gather}
\partial_t \langle \rho \rangle = \frac{\overline{D}}{4} \nabla^2 \langle \rho \rangle + \frac{\overline{D}\overline{\varepsilon}}{4}  \nabla \cdot \left( \nabla \cdot Q \right) - \overline{D}\overline{\varepsilon} \nabla \cdot \langle {\bf m} \rangle,\\
\partial_t \langle {\bf m} \rangle = \frac{\overline{D}}{4} \nabla^2 \langle {\bf m} \rangle + \frac{\overline{D}\overline{\varepsilon}}{8}  \begin{pmatrix}
\partial_{xx} - \partial_{yy} & 2\partial_{xy} \\
2\partial_{xy} & -\partial_{xx} + \partial_{yy}
\end{pmatrix} \langle {\bf m} \rangle
- \frac{\overline{D}\overline{\varepsilon}}{2} \left( \nabla \langle \rho \rangle + \nabla \cdot Q \right) \nonumber \\
+ \frac{q\gamma}{q-1} \left[ (\beta J - 1 -r \langle \rho \rangle^\alpha) - \frac{1}{2} \left(\frac{\beta J}{2\langle \rho \rangle} \right)^2 (7-3\beta J) \langle {\bf m} \rangle^2 \right] \langle {\bf m} \rangle,\\
Q = \frac{\beta J}{2\langle \rho \rangle} \begin{pmatrix}
\langle m_x \rangle^2 - \langle m_y \rangle^2 & 2\langle m_x \rangle \langle m_y \rangle\\
2\langle m_x \rangle \langle m_y \rangle & - \langle m_x \rangle^2 + \langle m_y \rangle^2
\end{pmatrix},
\end{gather}
with $\alpha = \xi -2$ and where $r$ is a constant which usually depends on $\beta$. These equations are reported in main text as Eqs.~(3)-(5) where the stochastic average $\langle \cdots \rangle$ have been omitted. When $r=0$, the mean-field equations are recovered, and we can take $r=1$ without any loss of generality (up to a rescaling of the density and the magnetization). We may note that
\begin{equation}
\nabla \cdot Q = - \frac{\nabla \langle \rho \rangle}{\langle \rho \rangle} \cdot Q + \frac{\beta J}{\langle \rho \rangle} \left[ (\langle {\bf m} \rangle \cdot \nabla) \langle {\bf m} \rangle - (\langle {\bf m}_\perp \rangle \cdot \nabla) \langle {\bf m}_\perp \rangle\right]
\end{equation}
with ${\bf m} = (m_x,m_y)$ and ${\bf m_\perp} = (-m_y,m_x)$. 

\section{Algorithm to obtain the time-averaged profiles}
Here, we would like to briefly discuss the algorithm that has been used to obtain the time-averaged density profiles in Fig. 2(c). For each instantaneous density profile $k$ at time $t$, we first move the center of each of the $n_b$ stripes ($n_b>1$) to a fixed point $x_0$ on the $x$-axis by doing a coordinate shifting (we always consider $x_0=L_x/2$) and perform an averaging over these stripes at time $t$. We then denote the averaged density over these stripes by $\bar{\rho}_k(x)$. We repeat this procedure for $n_p \simeq 200-300$ such instantaneous profiles and finally perform a thermal averaging over $n_p$ number of $\bar{\rho}_k(x)$ to obtain the time-averaged density profile as $\langle \bar{\rho}(x) \rangle = \frac{1}{n_p} \sum_{k=1}^{n_p}  \bar{\rho}_k(x)$.

\section{Order parameter distribution as a function of $\rho_0$ for $q=7$}

\begin{figure}[H]
\centering
\includegraphics[width=0.8\columnwidth]{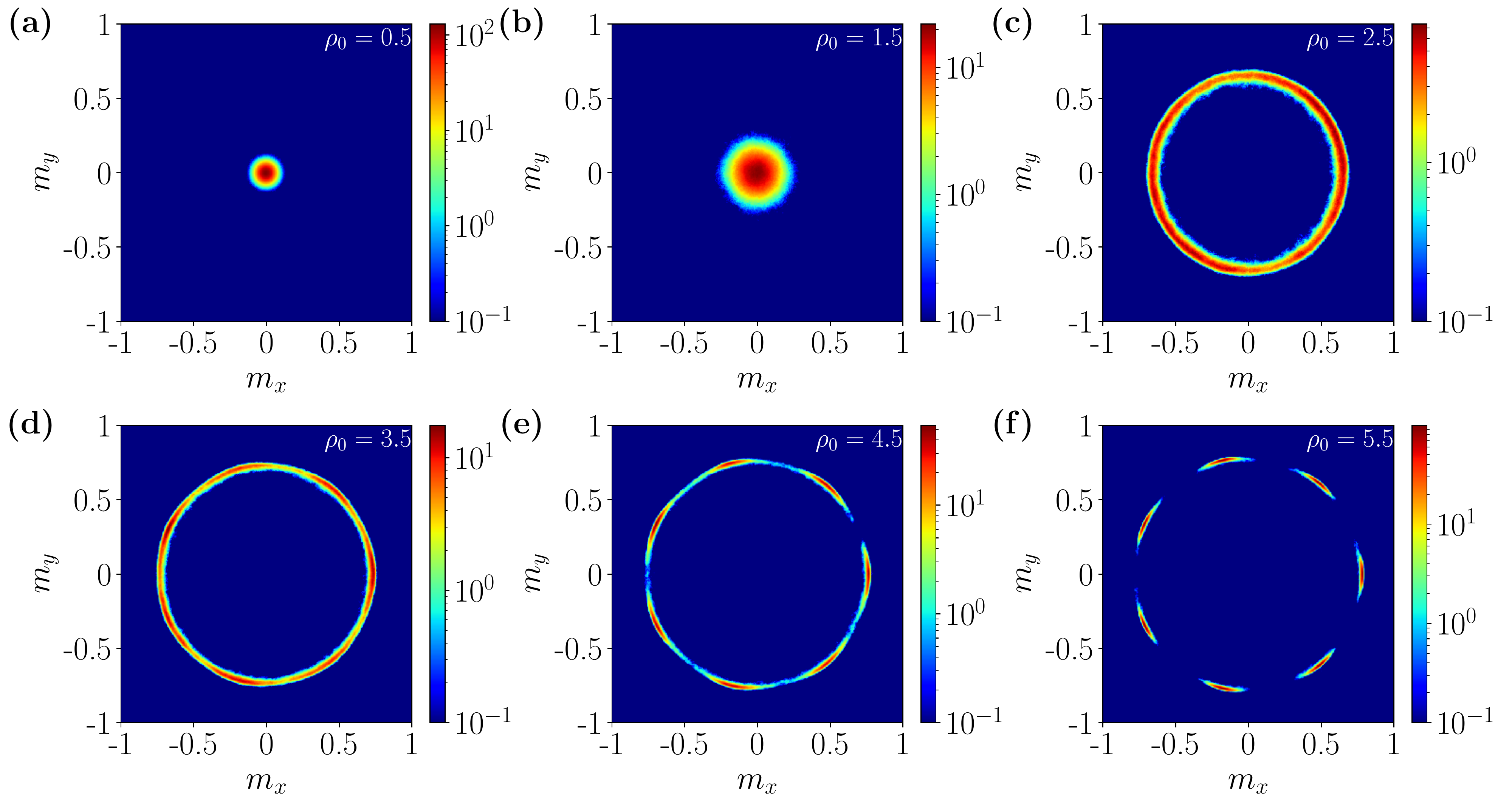}
\caption{(color online) Order parameter distributions for $q=7$ and several $\rho_0$ values. Parameters: $L=50$, $\beta=2$ and $\bar{\varepsilon}=0$.} 
\label{fig7}
\end{figure}

Distribution of the order parameter for a fixed $q$ and several densities are shown in Fig.~\ref{fig7} for $q=7$, $\beta=2$ and $\bar{\varepsilon}=0$ where we observe all the three different phases as a function of $\rho_0$: (a--b) homogeneous disordered phase with uniform distribution of spins at small $\rho_0$, where every spin points to a random direction, (c)--(e) a QLRO phase at intermediate densities with a ringlike distribution of the order parameter and (f) a LRO phase at a sufficiently high density where seven distinct spots correspond to the seven possible ordering states. One can notice that the spread of the distribution around the angles allowed for the clock spins in the LRO phase of $q=7$ is greater compared to $q=4$ and $q=5$ in Fig.~4 and things like the system size or the higher degeneracy of the state (due to which a perfect LRO phase like Fig.~4(a)--(b) might only be possible at a larger density) might be responsible for this. The conclusion which we draw from Fig.~\ref{fig7} is that the liquid phase for discrete $q$ values and for $\bar{\varepsilon}=0$ shows both QLRO and LRO at different parameter regimes, QLRO at smaller $\rho_0$ [$\rho_0>\rho^* (q)$] and LRO at a larger $\rho_0$. As shown in the context of equilibrium q-state clock models~\cite{clock-swarnajit}, one can also expect such a scenario as a function of temperature $\beta$.

\section{Effective exponents as a function of $q$}

\begin{figure}[H]
\centering
\includegraphics[width=0.8\columnwidth]{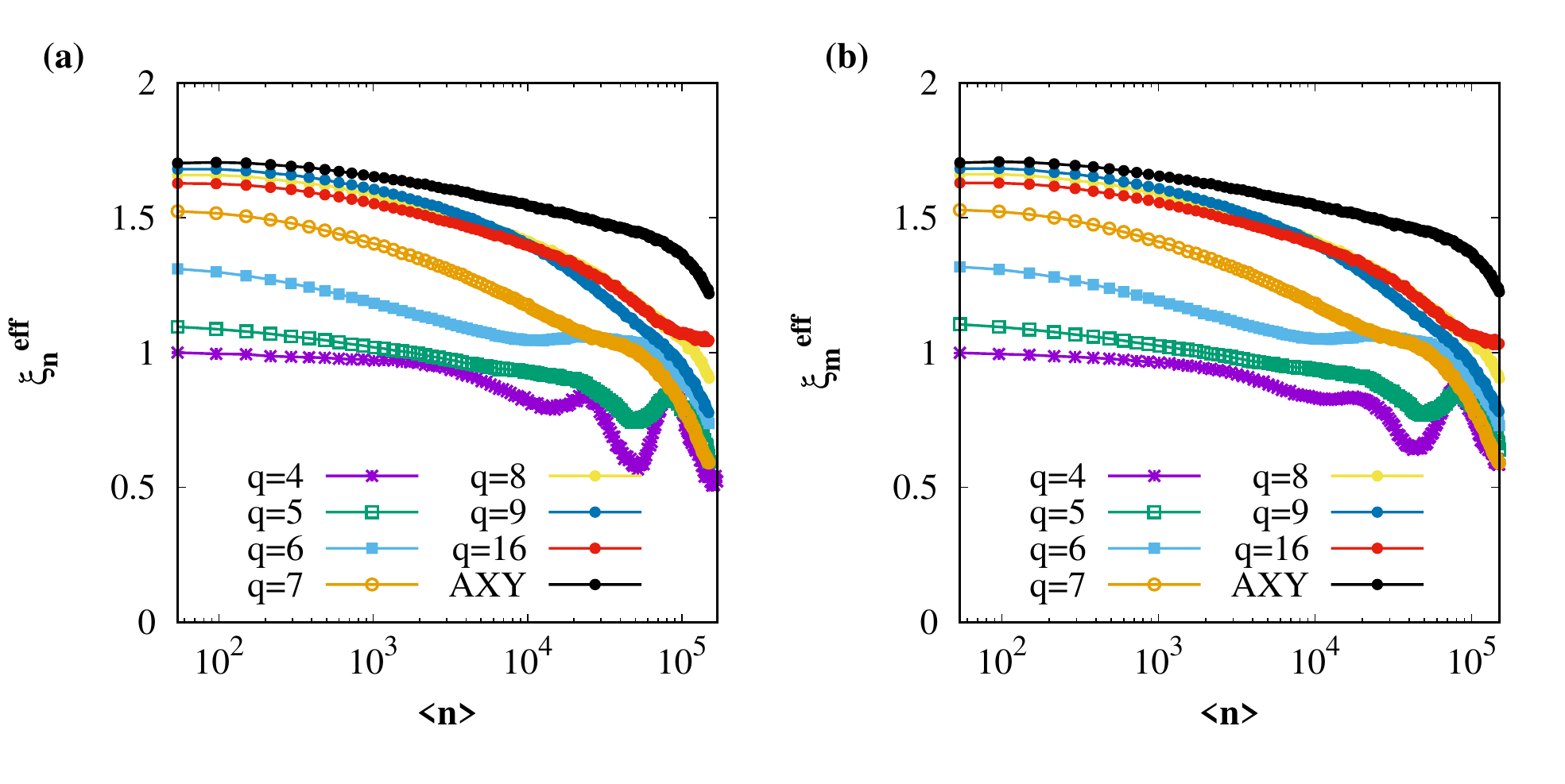}
\caption{(color online) (a)--(b) Effective exponents $\xi_n^{\rm eff}$ and $\xi_m^{\rm eff}$ versus $\langle n \rangle$ for the data plotted in Fig.~4(a)-(b). Parameters: $\beta=2$, $\bar{\varepsilon}=0.9$, and $\rho_0=6$.} 
\label{fig8}
\end{figure}

In Fig.~\ref{fig8}(a) and Fig.~\ref{fig8}(b), we respectively present the effective exponents $\xi_n^{\rm eff}=d[\ln(\Delta n^2)]/d[\ln \langle n \rangle]$ and $\xi_m^{\rm eff}=d[\ln(\Delta m^2)]/d[\ln \langle n \rangle]$ versus average particle number $\langle n \rangle$ (where $n$ denotes the particle number in subsystems of linear size $\ell$) for several $q$ values corresponding to the number fluctuations and magnetization fluctuations shown in Fig.~4(a)-(b). The exponents in Table~1 have been obtained by fitting the data in Fig. 4(a)–(b) to a power-law and since the extracted exponents depend on the interval along the x-axis to which the fits are restricted, a look at the log-log slope of the data or the effective exponent reveals more insight. The plots show ``plateaus'' around the extracted exponents for the corresponding $q$ values (see Table~1) for at most one decade of $\langle n \rangle$ and then $\xi^{\rm eff}$ decreases with increasing $\langle n \rangle$ when $\langle n \rangle$ approaches the total number of particles in the system and becomes smaller than 1 due to the finite-size cut-off at $\langle n \rangle=N=\rho_0L^2$, where $\Delta n^2$ vanishes.

\end{document}